\documentclass[twocolumn,superscriptaddress,english,prl,showpacs,longbibliography]{revtex4-1}
\usepackage[colorlinks=true,urlcolor=blue,citecolor=blue,linkcolor=blue]{hyperref} 

\usepackage{amsmath}
\usepackage{graphicx}
\usepackage{bm}
\usepackage{hyperref}
\usepackage{color}
\usepackage{amssymb}
\usepackage{graphicx}
\usepackage{color}
\usepackage{mathrsfs}
\usepackage{float}
\usepackage{indentfirst}
\usepackage{txfonts}
\usepackage{algorithm}
\usepackage{algpseudocode}
\usepackage{balance}

\newcommand{\s}{\mathbf {s}}

\newcommand{\vhead}{\mathbf {v}_\mathrm{head}}

\newcommand{\ghead}{\widehat{\mathcal G}_{\mathrm{head}}}
\newcommand{\gghead}{\widehat{\mathcal G}_{\mathrm{head}}}

\newcommand{\gtail}{\widehat{\mathcal G}_{\mathrm{tail}}}
\newcommand{\ggtail}{\widehat{\mathcal G}_{\mathrm{tail}}}

\newcommand{\hpsi}{\widehat{\psi}}
\newcommand{\hp}{\widehat{P}}
\newcommand{\hg}{\widehat{\mathcal G}}
\newcommand{\g}{{\mathcal G}}
\newcommand{\sstate}{\{\hpsi_i^{\mu}\}}
\newcommand{\fsim}{\textrm{fSim}}

\begin{document}
	
	\title{Solving the Sampling Problem of the Sycamore Quantum Circuits}
	
	\author{Feng Pan}
	\affiliation{
		CAS Key Laboratory for Theoretical Physics, Institute of Theoretical Physics, Chinese Academy of Sciences, Beijing 100190, China
	}
	\affiliation{
		School of Physical Sciences, University of Chinese Academy of Sciences, Beijing 100049, China
	}
	\author{Keyang Chen}
	\affiliation{
		CAS Key Laboratory for Theoretical Physics, Institute of Theoretical Physics, Chinese Academy of Sciences, Beijing 100190, China
	}
	\affiliation{
		Yuanpei College, Peking University, Beijing 100871, China.
	}
	\author{Pan Zhang}
	\email{panzhang@itp.ac.cn}
	\affiliation{
		CAS Key Laboratory for Theoretical Physics, Institute of Theoretical Physics, Chinese Academy of Sciences, Beijing 100190, China
	}
	\affiliation{
		School of Fundamental Physics and Mathematical Sciences, Hangzhou Institute for Advanced Study, UCAS, Hangzhou 310024, China 
	}
	\affiliation{
		International Centre for Theoretical Physics Asia-Pacific, Beijing/Hangzhou, China
	}
	
	\begin{abstract}
		We study the problem of generating independent samples from the output distribution of Google's Sycamore quantum circuits with a target fidelity, which is believed to be beyond the reach of classical supercomputers and has been used to demonstrate quantum supremacy. We propose a method to classically solve this problem by contracting the corresponding tensor network just once, and is massively more efficient than existing methods in generating a large number of \textit{uncorrelated} samples with a target fidelity. 
		For the Sycamore quantum supremacy circuit with $53$ qubits and $20$ cycles, we have generated $1\times10^6$ \textit{uncorrelated} bitstrings $\mathbf s$ which are sampled from a distribution $\widehat P(\mathbf s)=|\widehat \psi(\mathbf s)|^2$, where the approximate state $\widehat \psi$ has fidelity $F\approx 0.0037$. The whole computation has cost about $15$ h on a computational cluster with $512$ GPUs. 
		The obtained $1\times10^6$ samples, the contraction code and contraction order are made public.
		If our algorithm could be implemented with high efficiency on a modern supercomputer with ExaFLOPS performance, we estimate that ideally, the simulation would cost a few dozens of seconds, which is faster than Google's quantum hardware.
	\end{abstract}
	\maketitle
	The sampling problem of quantum circuits has been proposed recently as a specific computational task to demonstrate whether programmable quantum devices can surpass the ability of classical computations, also known as \textit{quantum supremacy} (or quantum advantage)\cite{arute2019quantum,aaronson2016complexity,bouland2019complexity,movassagh2019quantum,boixo2017simulation,aaronson2019classical,zlokapa2020boundaries,boixo2018characterizing, wu2021strong, zhu2021quantum}. As a milestone, in 2019, Google released the Sycamore quantum circuits to realize this approach for the first time~\cite{arute2019quantum}. The Sycamore quantum supremacy circuits contain $53$ qubits and $20$ cycles of unitary operations. Google has demonstrated that the noisy sampling task with fidelity $f\approx 0.002$ can be achieved experimentally using the quantum hardware in about $200$ sec, while they estimated that it would take $10,000$ yr on modern supercomputers.
	
	However, the computational time estimated by Google relies on a specific classical algorithm, the Schr\"odinger-Feynman algorithm \cite{aaronson2016complexity,markov2018quantum,arute2019quantum}, rather than a theoretical bound that applies to all possible algorithms. So, in principle, there could exist algorithms that perform much better than the algorithm used by Google, rejecting the quantum supremacy claim.
	Indeed, in this Letter, we provide such an algorithm based on the tensor network method.
	
	There have been great efforts to develop more efficient classical simulation algorithms. IBM has estimated that the $53$-qubit state vector of the Sycamore circuits can be stored and evolved if one could employ all the RAM and hard disks of the Summit supercomputer. However, it is apparently unrealistic to do such a numerical experiment. Recently a variety of methods have been proposed for this problem based on computing a single amplitude or a batch of amplitudes~\cite{boixo2017simulation,chen2018classical,guo2019general,gray2021hyper,huang2020classical} using tensor network contractions. In particular, \cite{huang2020classical} proposed contracting the corresponding tensor network $2000$ times to obtain $2000$ batches of amplitudes (each batch contains $64$ correlated bitstrings), then sample $2000$ perfect samples from the batches and mix them with 998000 random bitstrings to obtain samples with linear cross entropy benchmark (XEB) around $0.002$. However the computational cost of 
	%contracting the tensor networks for $2000$ times 
	such simulation is still too large, and the experiment has not been realized yet.
	
	Another attempt to pass the XEB test on the Sycamore quantum supremacy circuits is the recently proposed \textit{big-head} approach~\cite{pan_simulation_2022}, which can obtain a large number of correlated samples. Using $60$ GPUs for $5$ days, the authors of~\cite{pan_simulation_2022} generated $1\times10^6$ correlated samples with XEB $0.739$, passed the XEB test. We also noticed that very recent works~\cite{fu2021closing,liu2021redefining} implemented this approach on a supercomputer, and heavily reduced the running time for obtaining a batch of correlated samples. However, if the target of the simulation is not only passing the XEB test but also satisfying the constraint of obtaining \textit{uncorrelated} samples, as in the Sycamore experiments, then one needs to repeat the contraction thousands of times, making the computation cost unaffordable in practice. Moreover, a recent work~\cite{gao2021limitations} studied a particular method for obtaining high (average) XEB values but low fidelity, illustrating limitations of XEB as a measure for fidelity.
	
\begin{figure}[htb]
		\centering
		\includegraphics[width=0.95\columnwidth]{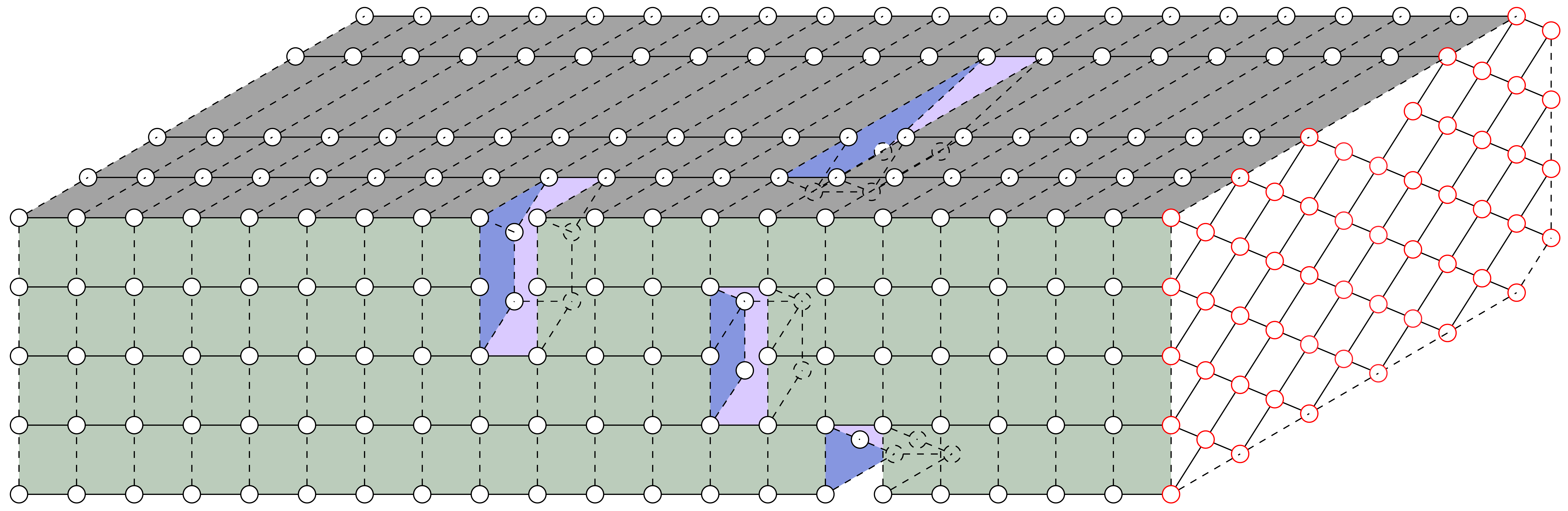}
		\caption{Pictorial representation of the three-dimensional tensor network corresponding to the Sycamore quantum circuit with $n=53$ qubits and $m=20$ cycles. There are $4$ holes in the tensor network designed for reducing the contraction complexity. Each hole is created by breaking two edges in a selected two-qubit gate and the companion edges, i.e. removing the entire two-qubit gate, as described in the main text.
			The result of contracting the three-dimensional tensor network using the sparse state method is $L=2^{20}$ groups of amplitudes, each group contains $l=2^6=64$ correlated bitstring amplitudes. That is, we have computed approximate amplitudes for $2^{26}=67,108,864$ bitstrings and finally sampled $2^{20}$ uncorrelated bitstrings from them.
			\label{fig:tn}}
	\end{figure}

	In this article, we propose a tensor network approach to solve the uncorrelated sampling problem for the Sycamore quantum supremacy circuits. Our method is based on contractions of the three-dimensional tensor network $\hg$ (Fig.~\ref{fig:tn}) converted from the quantum circuit. 
	A single contraction of $\hg$ produces $\sstate$ with $i=1,2,\cdots L$ and $\mu=1,2,\cdots l$, representing amplitudes of $L$ (randomly chosen) uncorrelated groups of bitstrings with each group containing $l$ correlated bitstrings. Since $\sstate$ contains a small portion of entries of a approximate state $\hpsi$ with fidelity $F$, we term it as a \textit{sparse state}. Based on the sparse state, we do importance sampling to obtain one sample from a group, finally generating $L$ uncorrelated samples from the approximate probability $\hp=|\hpsi|^2$, i.e., $L$ approximate samples from the output distribution of the quantum circuit with fidelity $F$.
	
	Our algorithm is massively more efficient than existing algorithms in generating a large number of uncorrelated samples.
	On the Sycamore circuits with $n=53$ qubits and $m=20$ cycles, we have successfully generated $L=2^{20}$ approximate samples with fidelity $F\approx 0.0037$ in about $15$ h using $512$ GPUs.
	We remark that to the best of our knowledge this is the first time that the sampling problem of the Sycamore quantum supremacy circuits (with fidelity larger than Google's hardware samples) with $n=53$ qubits and $m=20$ cycles is solved in practice classically.
	
	\paragraph{Method. --- }
	The quantum circuits $U$ can be regarded as a unitary tensor network $\mathcal G$ with matrices (corresponding to single-qubit gates) and four-way tensors (corresponding to two-qubit gates) connecting to each other.
	For the Sycamore circuits where the qubits are placed on a two-dimensional layout, the corresponding $\g$ is a three-dimensional tensor network as illustrated in Fig.~\ref{fig:tn}.
	The initial state (the leftmost layer) and the final state (the rightmost layer) act as two boundary conditions to $\mathcal G$.
	The initial state is always a product state so acts as a set of vectors; while the final state is represented as either a giant tensor or a set of small tensors (including vectors) depending on how many amplitudes we request in contraction of $\mathcal G$.
	
	If we request all amplitudes of the final state, the final state acts as a giant tensor with size $2^n$, which requires a storage space exponential to the number of qubits. If we request only one amplitude of the final state, then the boundary is a product state and acts as a set of vectors. Another case considered in the literature is the batch contraction~\cite{huang2020classical,pan_simulation_2022}, which requests amplitudes for $l$ correlated bitstrings and gives a tensor with size $l$ as the final boundary condition for $\mathcal G$.
	In this work our target is different: we request a large number of amplitudes for uncorrelated bitstrings, from single contraction of $\hg$, a slightly perturbed version of $\mathcal G$.
	
	Tensor network $\hg$ is created by breaking (removing) $K$ edges (connections) in $\g$. The edge breaking is implemented by inserting 
	$E=\left(\begin{matrix}1\\0\end{matrix}\right)\otimes \left(\begin{matrix}1\\0\end{matrix}\right)$
	in between the two tensors that the edge is connecting.
	In this work, we select $K$ edges from input indices of $K/2$ two-qubit gates. Pictorially it represents as drilling ${K}/{2}$ holes in the three-dimensional graphical representation of $\hg$ as shown in Fig.~\ref{fig:tn}.
	The position of holes are determined such that contracting $\hg$ is much easier than contracting $\g$, but with the price of decreasing the fidelity.
	The amount of decreased fidelity can be estimated using the expression of $E$ as a specific Pauli error matrix $ 
	E=\frac{1}{2}I+\frac{1}{2}\sigma_z$,
	with $I=\left(\begin{matrix}1&0\\0&1\end{matrix}\right)$ and $\sigma_z=\left(\begin{matrix}1&0\\0&-1\end{matrix}\right)$. The effect of the edge breaking can be understood as breaking the system into a summation of two subnetworks. The first subnetwork is a copy of the original one which preserves the information of the original final state, while the second subnetwork with the action of $\sigma_z$ completely destroys the information of the original final state.
	Since the weight of each subnetwork is $1/2$~\cite{markov2018quantum}, one then estimates that each edge breaking decreases the fidelity $F$ by a factor of $1/2$.
	After breaking $K$ edges in $\g$, we arrive at $\hg$. If we contract $\hg$ and obtain a full amplitude state vector $\hpsi$, it would be an approximation to the final state $\psi$ of $\g$, with fidelity estimated as $F_K\approx 2^{-K}$.
	
	The simulation method based on tensor network contractions can be regarded as Feynman's path-integral approach, because the tensor contractions effectively sum over an exponential number of paths which are considered to be orthogonal to each other hence contributing equally to the obtained amplitudes.
	Under this viewpoint, the hole drilling in $\mathcal G$ can be understood as omitting some paths in the path-integral approach, summing over only a fraction of $2^{-K}$ paths, giving fidelity $F_K\approx 2^{-K}$. 
	
	In this work we only request the sparse state, the amplitudes for $L\times l$ bitstrings which are grouped into $L$ groups with each group containing $l$ correlated bitstring amplitudes (in the practical $L=2^{20}$ and $l=2^6$). They are given according to a generation process in advance and kept fixed during the contraction. 
	
	However, contracting $\hg$ to arrive at the $L\times l$ size sparse state is a very difficult task, and the space complexity of the contraction would be much larger than $L\times l$. To solve the problem we extend the \textit{big-head} algorithm proposed in~\cite{pan_simulation_2022}. In the big-head algorithm, the three-dimensional tensor network is cut into two parts, $\gghead$ whose contraction cost dominates the whole computation, and $\ggtail$ which contains all the qubits in the final state and can completely reuse the contraction results of $\gghead$ for computing all the requested amplitudes. In this work, the big-head method is extended to work with the sparse state (rather than a batch of correlated bitstrings in~\cite{pan_simulation_2022}). To this end, we need to balance the computation cost of $\ghead$ and the cost of $\gtail$. The contraction results of $\ghead$ is a vector $\vhead$, with size much larger than our storage limit, so in practice, we enumerate $k$ entries in $\vhead$, that is, making $2^k$ slices of the $\vhead$, each slice has size $2^{29}$. Given each slice of $\vhead$, the $\ghead$ is contracted with a good contraction order and local dynamic slicing, similar to~\cite{pan_simulation_2022}. 
	
	The boundary condition given by the sparse state is heavier to deal with than the boundary conditions of $\ghead$. We proposed a new \textit{zigzag} method for finding a good contraction order.  The method starts at the beginning boundary of $\gtail$, contracting neighboring tensors in a complexity-greedy manner all the way towards the boundary of the sparse state, then turns around to contract greedily the tensors and come back to the beginning boundary. The process is repeated until all the tensors in $\gtail$ are contracted, and the sparse state $\{\psi_i^{\mu}\}$ is obtained. The spirit of the zigzag contraction order is to make use of both boundaries to reduce the space and time complexity of contraction. For more details about the head-tail splitting of the circuits, the sparse state contraction method and slicing technique, please refer to the Supplemental Material~\cite{SI}.
	
	In the Sycamore circuits, two-qubit unitary transformations are parametrized using the fSim gates
	\begin{equation}
		\label{eq:fsim}
		\text{fSim}(\theta, \phi) = 
		\begin{bmatrix}
			1 & 0 & 0 & 0 \\
			0 & \cos{\theta} & -i\sin{\theta} & 0 \\
			0 & -i\sin{\theta} & \cos{\theta} & 0 \\
			0 & 0 & 0 & e^{-i\phi}
		\end{bmatrix}\;.
	\end{equation}
	Specifically, the parameters in Google's experiments~\cite{arute2019quantum} are tuned to $\theta \approx \pi / 2$ in order to keep the decomposition rank equal to $4$ with a near-flat spectrum, that is, the singular values of the $4\times 4$ matrix obtained by reshaping the fSim gate are almost identical~\cite{arute2019quantum}. This setting significantly increases the cost of classical simulations when compared with controlled-Z gates which has decompositional rank $2$, in exact simulations and in approximate simulations~\cite{zhou2020limits,pan2020contracting}. 
	
	However we observe that in our approach there are two situations that we can explore the low rank structures.\\ 
	(i) In the hole drilling, when the two input indices ($\alpha$ and $\beta$) of the $\fsim$ gate are cut, i.e. applying two Pauli errors gate as 
	$A = \left(\left[\begin{matrix}1&0\\0&0\end{matrix}\right]\otimes \left[\begin{matrix}1&0\\0&0\end{matrix}\right]\right)\cdot\fsim(\theta,\phi),$
	as illustrated in Fig.~\ref{fig:fsim} top.
	It evaluates to a rank-one matrix $B=\left[\begin{matrix}1&0\\0&0\end{matrix}\right]$, hence the fSim gate can be replaced by two $(1,0)$ vectors without decreasing fidelity.\\
	(ii) In enumerating $k$ entries of $\vhead$ as well as in the slicing process, fixing an index is regarded as breaking one input edge in the tensor diagram as illustrated in Fig.~\ref{fig:fsim} bottom (e.g. the top left edge $\gamma$ of tensor $D$ is cut), giving a three-way tensor $E$. Although the decompositional rank of $E$ on the bottom right index $\omega$ is $2$, the corresponding squared singular values, $[\sin^2(\theta)+1,\cos^2(\theta)]$, are heavily imbalanced in the Sycamore circuits with $\theta\approx {\pi/2}$. In this way we can do a rank-one approximation by dropping the singular vectors corresponding to the squared singular value $\cos^2(\theta)$. This rank-one approximation decreases the fidelity approximately by a factor $[\sin^2(\theta)+1]/2$, while effectively break another edge $\omega$, which we term as the \textit{companion edge} in the tensor network. For total $k$ 
	% enumerations in $\vhead$ and 
	selected slicing edges in the tensor network , we do the rank-one approximation for associated $\fsim$ gates, cutting $k$ associated companion edges. This decreases the fidelity $F$ by a factor $\prod_{i=1}^k[\sin^2(\theta_i)+1]/2$.
	%, while at the same time removes $k$ edges from the tensor network. 
	
	\begin{figure}[h]
		\centering
		\includegraphics[width=0.8\columnwidth]{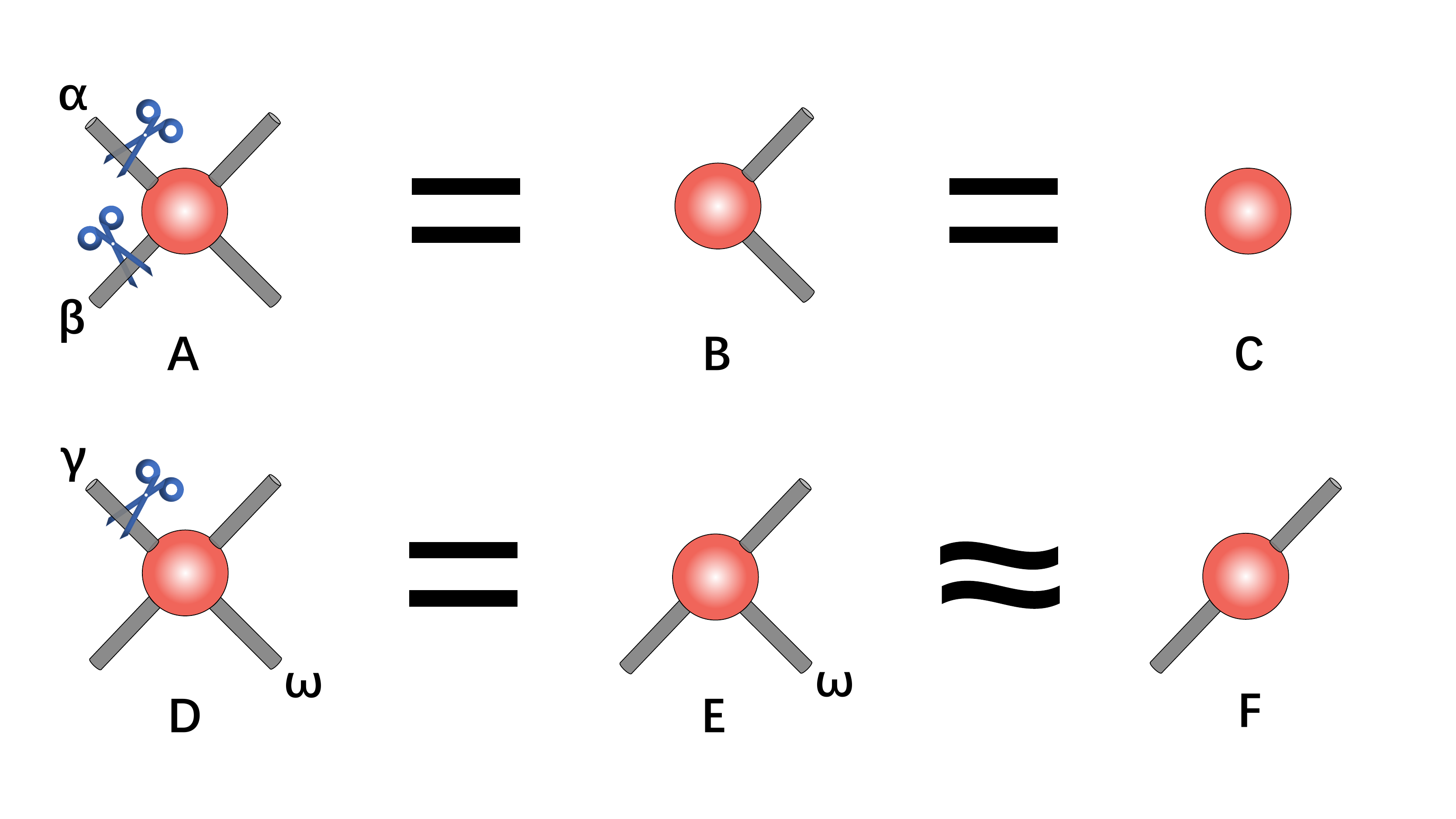}
		\caption{Two situations where we can explore the low-rank structures. (Top) When two indices $\alpha,\beta$ are pinned to $0$ for the fSim gate. The result is a rank-one matrix $B$ effectively equals to a scalar $c=1$ in the tensor network. (Bottom) When one index $\gamma$ of an fSim gate is pinned, the resulting tensor $E$ has decompositional rank $2$ but with imbalanced singular values $[\sqrt{\sin^2(\theta)+1},\cos(\theta)]$ with $\theta\approx \pi/2$.
			\label{fig:fsim}}
	\end{figure}
	
\begin{figure}[htb]
		\centering
		\includegraphics[width=0.98\columnwidth]{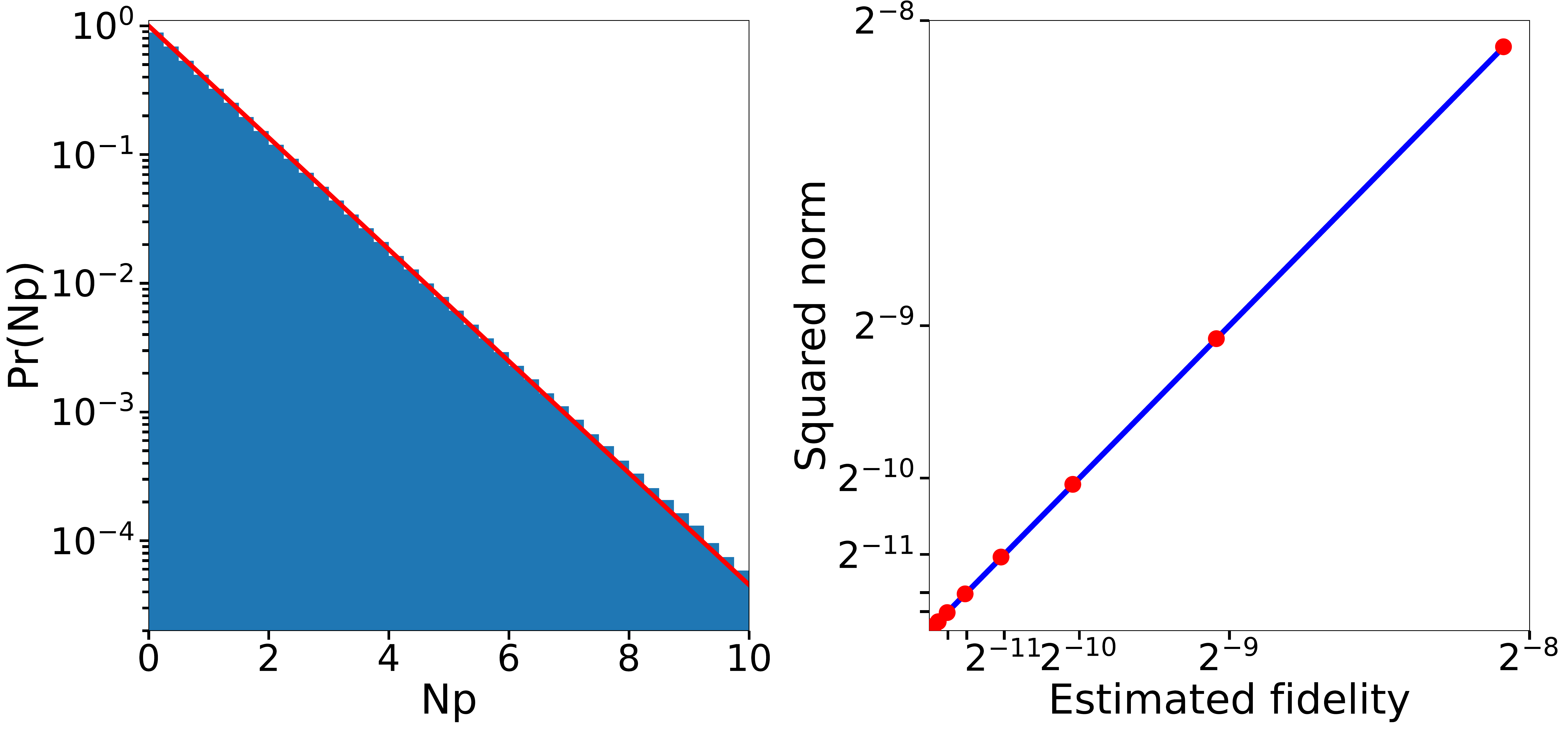}
		\caption{(Left:) Histogram of approximate bitstring probabilities $p(\s)=|\widehat \psi(\s)|^2/\mathcal{N}_s$ for $2^{26}$ bitstrings obtained from the Sycamore circuits with $n=53$ qubits and $m=20$ cycles. $\mathcal{N}_s$ is the norm factor and $N=2^n$. The estimated fidelity $\widehat\psi(\s)$ to the true final state $\psi(\s)$ is $F\approx 0.0037$. The red line denotes the Porter Thomas distribution. (Right:) Comparison between the estimated fidelity (blue lines) and the norm factor of $\widehat \psi_L(\s)$ obtained by summing over a fraction of paths. \label{fig:dis}}
	\end{figure}

	After contracting $\hg$ using the methods we have introduced above, the sparse state $\sstate$ is obtained, which are selected from $2^n$ entries of a state $\hpsi$, with fidelity to the true state estimated as
	$F_{\mathrm{estimate}}\approx 2^{-K}\prod_{i=1}^k[\sin^2(\theta_i)+1]/2$.
	Since the sparse state $\sstate$ is composed of $L$ groups and each group contains $l$ amplitudes, we use a Markov chain to sample one bitstring out of $l$ amplitudes in each group using the Metropolis algorithm~\cite{newman1999monte}, producing $L$ samples which is considered as unbiased samples from $\hpsi$. We also note that if $|\hpsi|^2$ follows the Porter-Thomas distribution~\cite{porter1956fluctuations,brody1981random,boixo2018characterizing} (as we verify empirically in Fig.~\ref{fig:dis}), we can use the frugal sampling~\cite{markov2018quantum,arute2019quantum} which is much faster and guaranteed to give near-perfect samples with $l=64$.
	We remark that to obtain $L$ uncorrelated samples, only groups need to be independently and randomly generated, it is not necessary to maintain uncorrelated bitstrings inside of each group~\cite{villalonga2019flexible}. The validations of our approximate sampling method using smaller Sycamore circuits can be found in the Supplemental Material~\cite{SI}.
	
	\paragraph{Results. ---}
	We focus on the Sycamore circuits with $n=53$ qubits, $m=20$ cycles, sequence \textit{ABCDCDAB}, which have been used to demonstrate the quantum supremacy based on the estimated $10,000$ yr for classical simulations~\cite{arute2019quantum}.
	We first simplify the tensor network by contracting order-one and order-two tensors into their neighbors, resulting in a tensor network with $n = 455$ tensors. To arrive at $\hg$, we chose $K=8$ edges to break, they are associated with $4$ $\fsim$ gates. Using the low-rank structure, we completely remove the two-qubit gates by introducing proper Pauli error gates. This gives $4$ holes marked in Fig.~\ref{fig:tn}. This approximation decreases the fidelity by a factor $2^{-8}$.
	Then the tensor network is divided into two parts, the head part $\ghead$ and the tail part $\gtail$.
	
	We introduce $6$ slicing edges in contracting $\ghead$. The space and time complexity are $2^{30}$ and $2.3816\times10^{13}$ respectively.
	Contracting the $\ghead$ results in a tensor $\vhead$ of size $2^{45}$, which we cannot store, so we enumerate $16$ entries of the $\vhead$, creating $2^{16}$ subtasks of tensor network contraction, each of which corresponds to a configuration of $16$ binary variables. 
	
	In each subtask, $\vhead$ is sliced to a tensor with size $2^{29}$, which works as a boundary for $\gtail$. For the Sycamore circuits with $n=53$ qubits and $m=20$ cycles, we set $L=2^{20}$ and $l=2^6$, i.e. organizing the requested bitstrings to $2^{20}$ independent groups, each of which contains $2^6$ bitstrings. It acts as another boundary of $\gtail$. In contracting $\gtail$, we introduces $7$ local slicing edges, and the space and time complexity in our sparse state contraction scheme are $2^{30}$ and $2.9425\times10^{13}$ respectively. The overall time complexity of the entire computation (for finishing $2^{16}$ subtasks) is $3.489\times10^{18}$, which is slightly lower than the previous work~\cite{pan_simulation_2022} in computing a large batch of correlated bitstring amplitudes, and~\cite{huang2020classical} in computing a small batch of correlated bitstrings. 
	
	In contracting $\gtail$, there are $5$ slicing edges associated with a companion edge. Together with the $16$ companion edges in enumerating $\vhead$, there are totally $k=21$ companion edges.
	We do further low-rank approximations on the $k=21$ associated $\fsim$ gates, decreasing the fidelity by a factor $\prod_{i=1}^{21}[\sin^2(\theta_i)+1]/2\approx 0.9565, $ where $\theta_i$ in the equation denotes the parameters of involved $\fsim$ gates.
	Together with the fidelity decreasing introduced in hole drilling, the final fidelity is estimated as
	\begin{equation}
		F_{\mathrm{estimate}}= 2^{-8}\times 0.9565\approx 0.0037. 
	\end{equation}
	
	%The head-tail splitting of the circuits as revealed in the 2-dimensional qubit layout, and the actual contraction order are provided explicitly in Supplemantery Materials.
	To increase the GPU efficiency, the branch merge strategy \cite{huang_efficient_2021, pan_simulation_2022} was adopted during the contraction. 
	After branch merging, the GPU efficiency is $31.76\%$ for $\ghead$ and $14.27\%$ for $\gtail$, the overall efficiency is $18.85\%$. 
	%The detailed data about the complexity, estimated fidelity, and GPU efficiency are listed in Supplemantery Materials.
	We use the \textit{Complex64} as data type in contraction. The contraction time of $\ghead$ for one subtask is around 112 sec and that of $\gtail$ is around 315 sec, summing to 427 sec for completing a single subtask. The entire simulation with $2^{16}$ subtasks is finished in about 15 h using a computational cluster with 512 GPUs.
	Detailed data about the complexity, estimated fidelity, GPU efficiency are listed in the Supplemental Material~\cite{SI}.
	
	By summing over $2^{16}$ paths, $\hg$ is contracted. The results are $2^{26}$ bitstrings amplitudes grouped into $2^{20}$ uncorrelated groups corresponding to partial bitstrings $\mathbf x\in \{1,0\}^{47}$ that are uniformly and randomly selected. Each group is composed of $2^6=64$ correlated bitstrings corresponding to $6$ open qubits. As a sanity check, we compute the squared norm $\mathcal N=  \sum_{i=1}^{2^{20}}\sum_{\mu=1}^{64}|\hpsi_i^\mu|^2$ of the sparse state by summing only a fraction of total paths, and compare to the expected fidelity with partial summation (i.e. the fraction of the paths). The result are shown in Fig.~\ref{fig:dis} right, where we can see that they coincide to each other.
	Using the norm of the sparse state we can estimate the normalization factor of the approximate distribution as $2^{27}\mathcal N$, and compute the approximate probability of bitstrings. The histogram of the probability is plotted in Fig.~\ref{fig:dis} left, where we can see that it fits very well to the Porter-Thomas distribution.
	
	Finally we generate $2^{20}$ uncorrelated bitstrings from the distribution of the sparse state using the MCMC importance sampling. The other method that we have tried is the frugal sampling which is guaranteed to work well~\cite{markov2018quantum,arute2019quantum} as the distribution fits to the Porter-Thomas distribution.
	
	\paragraph{Discussions. ---}
	We have presented a tensor network method for solving the approximate (uncorrelated) sampling problem of the Sycamore quantum circuits which was thought to be impossible for classical computations. 
	Using our algorithm the simulation for the Sycamore circuits with $n=53$ qubits and $m=20$ cycles is completed in about $15$ hours using $512$ V100 GPUs. 
	There are several places that the proposed algorithms can be further speed up. First, our contraction algorithm is straightforwardly implemented using Pytorch. We expect that using a library that is more suitable for tensor contractions, such as the cuQuantum~\cite{cuquantum}, the computational efficiency can be greatly increased. Second, in recent days a modern supercomputer could achieve a performance of ExaFLOPS ($10^{18}$ floating-point operations per second). If our simulation of the quantum supremacy circuits (with about $2.79\times 10^{19}$ floating-point operations without branch merging) can be implemented in a modern supercomputer with high efficiency, in principle, the overall simulation time can be reduced to a few dozens of seconds, which is faster than Google's hardware experiments.
	
	\begin{acknowledgments}
		The Sycamore circuit files are retrieved from~\cite{data}, and the circuits are loaded with Cirq~\cite{cirq_developers_2021_5182845} script contained in the data repository and converted to the tensor network $\g$. 
		Our contraction code is implemented using \textit{Pytorch} (version 1.7.2) with \textit{cudatoolkit} (version 10.1).
		The samples and the contraction code together with the contraction orders and slicing indices for reproducing our results are available at~\cite{code}. 
		The computation was carried out at the Cloud Brain I Computing Facility at the Peng Cheng Laboratory and HPC cluster of ITP, CAS.
		We acknowledge Pengxiang Xu for help in performing computations on the Cloud Brain computers. 
		We also thank Xun Gao, Ying Li, Lei Wang, Song Cheng, and Xiao Yuan for helpful discussions. P.Z. was supported by Chinese Academy of Sciences Grant No. QYZDB-SSW-SYS032, and Projects No. 11747601 and No. 11975294 of National Natural Science Foundation of China.
	\end{acknowledgments}

\bibliography{ref.bib}
\onecolumngrid
\appendix
\section{The sparse-state method}
In the main text, we have introduced the sparse-state tensor contraction method to calculate amplitudes of $2^{20}$ uncorrelated bitstrings in just one tensor network contraction. Here we give a detailed description of how this method works.

First we introduce the details of generating $L$ independent samples using $L\times l$ bitstrings. We draw $L$ groups uniformly and randomly from $2^{n}/l$ entries. In the simulation of the Sycamore circuit with $n = 53$ qubits, we specifically set $l=2^6$, meaning that $6$ qubits are chosen as totally open (all the configurations belonging to the subspace appear in the samples). We chose $2^6$ bitstrings for each group because when the final states follows the Porter-Thomas distribution in a random circuit distribution, sampling methods e.g. the frugal sampling~\cite{markov2018quantum,arute2019quantum} guarantees to give near-perfect samples with $l=64$.
 
Let us denote a sample using a bitstring $\mathbf s\in \{0,1\}^{53}$, 
            and write the bitstring as a concatenation of two sub-bitstrings $\s =\{\s_1;\s_2\}$, with $\s_1\in\{0,1\}^{47}$ and $\s_2\in\{0,1\}^{6}$. In detail, we generate $L$ sub-bitstrings $\{\s_1\}$ uniformly and randomly out of $2^{47}$ possible ones by choosing each of its binary entry by flipping a coin. Then, each $\s_1$ is concatenated to all $2^6$ $\s_2$ sub-bitstrings to create a group of size $64$, and finally, we arrive at $L\times l$ samples.

\begin{figure*}[htb]
	\centering
	\includegraphics[width=\columnwidth]{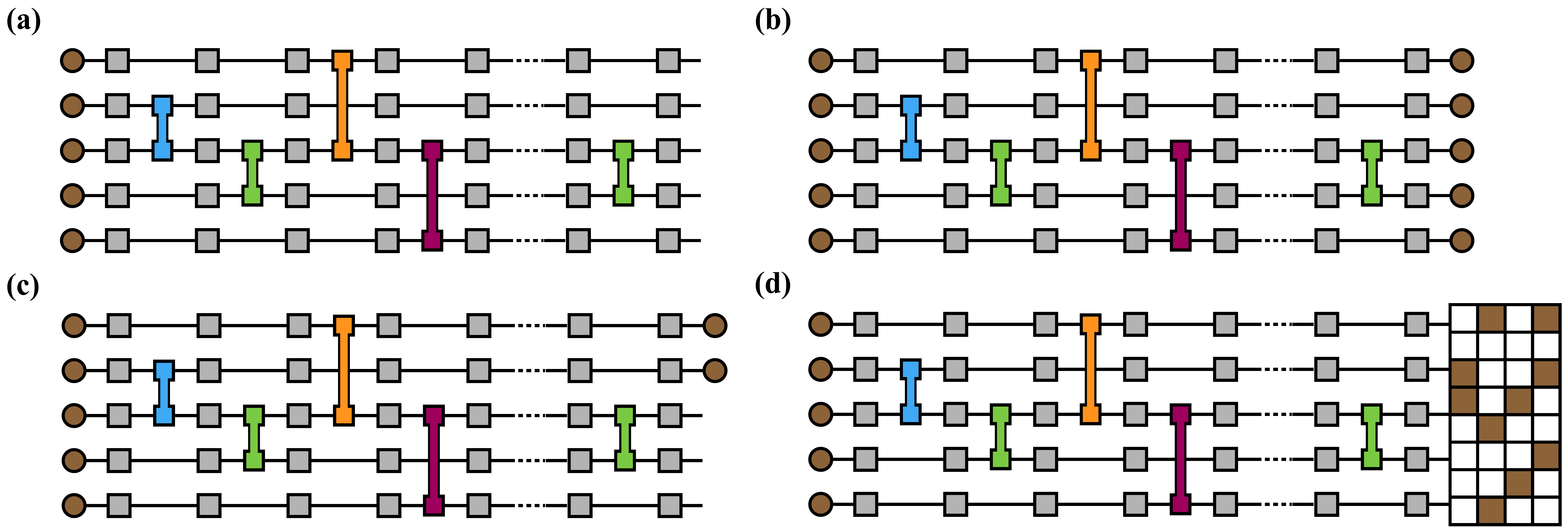}
	\caption{Four different types of quantum circuit simulation strategies by contracting the corresponding tensor network. (a) Full amplitude simulation. (b) Single amplitude simulation. (c) Batch simulation. (d) Sparse-state simulation. \label{fig:circuit_contrction}}
\end{figure*}

Fig.~\ref{fig:circuit_contrction} illustrates different types of quantum circuit simulation strategies by contracting the corresponding tensor network. Intuitively, the full amplitude simulation (type (a)) results to all amplitudes of the Hilbert space $U|0\rangle$. However, contracting such a tensor network requires $2^{n}$ space complexity which is beyond the reach of current computational devices when qubit number $n$ larger than 50. In order to alleviate the exponentially large storage requirement, the Schrodinger-Feynman algorithm~\cite{markov2018quantum, arute2019quantum} was introduced. Unfortunately, this method dramatically increases the computational cost.

To mimic the actual measurement procedure in the quantum experiment, the single amplitude simulation (type (b)) was considered~\cite{markov2008simulating, gray2020hyper}, which projects all qubits onto a basis vector in the computational basis $| x \rangle$ after execution of the quantum gates. The contraction result of the tensor network is $\langle x|U|0\rangle$, i.e. a single amplitude. In order to calculate all the amplitudes of $m$ bitstrings sampled from the quantum experiments, the contraction has to be repeated $m$ times. The contraction cost of the single amplitude simulation will be relatively small since the corresponding tensor network is closed. But if the repetition time $m$ is large, the overall cost would be unacceptable.  

To get a bunch of bitstring amplitudes in one contraction (type (c)), the so-called batch simulation was introduced. The spirit of the batch simulation is only projecting some of the qubits into a computational basis other than all of them. The result of the batch simulation is the amplitudes of all bitstrings in a Hilbert subspace on non-projected qubits, the entries of the projected qubits will be fixed according to the computational basis used. The batch simulation can be used for sampling~\cite{villalonga2019flexible, huang2020classical}, spoofing the XEB test~\cite{pan_simulation_2022} and full amplitude simulation~\cite{pan_simulation_2022}. However, to obtain uncorrelated samples one still needs to repeat the batch contraction a large number of times, making the overall computation not affordable for large circuits.

For the sampling problem of the quantum circuits studied in this paper, a more efficient way is to project the quantum state onto the set of basis vectors corresponding to the bitstring being sampled, we term it as the sparse-state of the entire Hilbert space, as illustrated in Fig~\ref{fig:circuit_contrction} (d). In this example, there are overall $2^5=32$ bitstrings in the Hilbert space (represented by the checkerboard), but only 10 bitstrings are sampled, corresponding to the brown entries. The sparse-state at the end of the quantum circuit will guide the contraction of the corresponding tensor network. So amplitudes of the sparse-state can be computed by contracting the tensor network for just once.

\begin{figure*}[htb]
	\centering
	\includegraphics[width=\columnwidth]{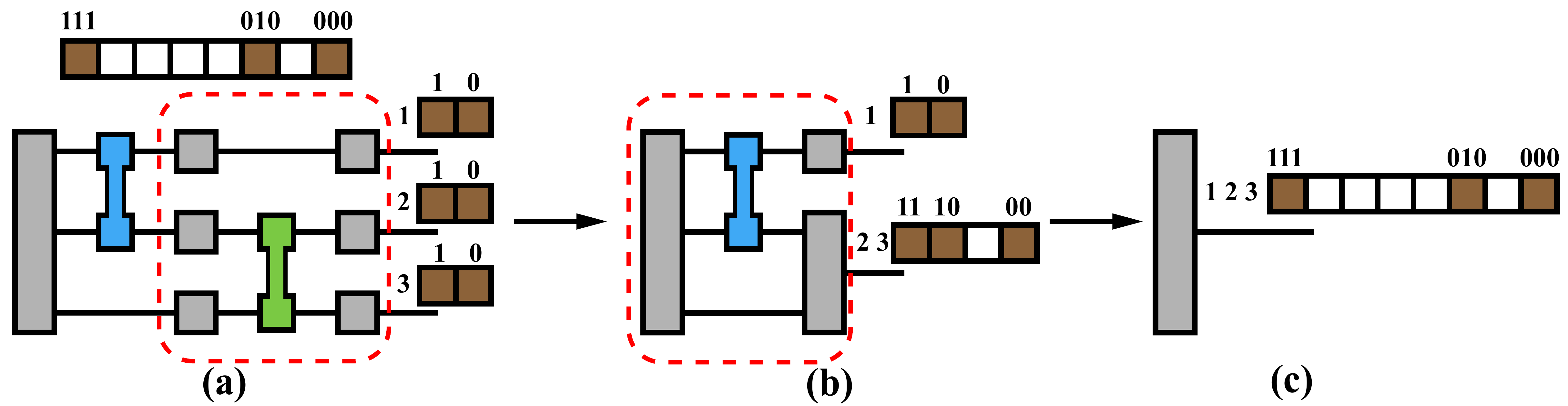}
	\caption{An example to illustrate the sparse-state simulation. \label{fig:sparsestate_con}}
\end{figure*}

In order to achieve such a goal, the sparse-state contraction scheme is introduced in the main text. Here we use a small example to illustrate how it works. Suppose we would like to simulate a quantum circuit with three qubits and obtain the amplitudes of $\{111,010,000\}$, as shown in Fig.~\ref{fig:sparsestate_con}(a) (here the front part of the circuit has been contracted into a big tensor lying at the left hand, and the brown part in the above checkerboard represents bitstrings we want to obtain). Initially, all the qubits at the final state will be "open". After contracting the tensors, within the red dashed line in (a), we will get the tensor network in (b). During the contraction, qubit 2 and qubit 3 are merged together. If we check the target bitstrings, we find that they have only three unique configurations at the qubit 2 and qubit 3 locations: $\{11,10,00\}$. Thus when qubit 2 and 3 merge, only three out of four entries need to be calculated, as illustrated by the checkerboard in Fig.~\ref{fig:sparsestate_con}(b). After that, the whole tensor network will be contracted together and all the three qubits will be merged together into a vector in the Fig.~\ref{fig:sparsestate_con}(c). When qubit 1 is merged with qubit 2 and 3, the target bitstrings having length 2 and length 3 merged into a length-3 vector, and the blank entries shown in Fig.~\ref{fig:sparsestate_con}(c) will not be calculated. After all the contractions, we obtain the final contraction result which is a vector whose entries represent the amplitudes of 3 target bitstrings.

From the above descriptions, we summarize that the key spirit of the sparse-state contraction is that when contracting two tensors involving qubits at the final state, one should always refer to the target bitstrings and find out which entries of the merged dimension are required to be calculated. And this is essentially the underlying information about the sparse-state boundary condition we put at the end of the tensor network associated with the quantum circuit. Since the sparse-state simulation method avoids all the unnecessary calculation of bitstrings except for the requested ones, it hugely reduces the overall complexity (especially the space complexity) of tensor network contraction corresponding to the random circuit sampling task, which makes the classical simulation of quantum supremacy circuits possible.

\section{Details about the numerical simulations}

The task of the simulation is to generate $2^{20}$ independent samples with XEB greater than Google's value $0.002$.
The sparse-state contraction scheme, which can incorporate all bitstring amplitudes calculations into one tensor network contraction, has been introduced in the main text and the previous section. The method can decrease the overall contraction complexity of obtaining amplitudes of multiple uncorrelated bitstrings. To solve the practical sampling problem there are still many problems left, such as the choice of bitstrings to be calculated and the exact computational cost to pass the quantum supremacy. In this section, we will discuss these problems in detail.

\subsection{Bitstrings choice and sampling details}

To pass the XEB test with XEB greater than $0.002$, at least one million bitstring samples should be obtained. 
Even though we have no information about the final state of the quantum circuit, we can compute a large number of bitstring amplitudes and probabilities and perform importance sampling from the probability distribution over the bitstrings. Suppose the distribution of bitstrings is $p(x)$ and $L$ bitstring samples are required, we can calculate $l\times L$ bitstring amplitudes and sample $L$ bitstrings. It has been shown in~\cite{markov2018quantum} that if $l=10$, the sampled distribution $\hat{p}(x)$ will be pretty close to the true distribution when the underlying distribution follows the Porter-Thomas distribution. The requirement is all $l\times L$ should be distinct and drawn uniformly at random. We notice that recently, a new sampling technique has been introduced to further decrease the sampling overhead $M$ to 2~\cite{kalachev_classical_2021}. 

For simulating quantum circuits by tensor network contraction method, the more friendly way to do sampling is the subspace sampling, which is introduced in~\cite{villalonga2019flexible} and also used by~\cite{huang2020classical}. The subspace sampling method does not require $l\times L$ distinct and uniformly random bitstrings, but $L$ groups of bitstrings, each group contains $l$ bitstrings coming from the same subspace. In this work we set $l=2^6=64$, for the sparse-state contraction of the Sycamore quantum circuit with $n=53$ qubits, which means that we choose 6 qubits to be totally open (all the configurations of that subspace will be in the samples), and project other 47 qubits into $L$ uniformly random bitstrings at $2^{47}$ Hilbert subspace. The computational overhead compare to projecting all qubits into $L$ uniformly random bitstrings can be negligible if the number of open qubits is not that large and their locations are carefully chosen.

In the practice of simulating Sycamore circuit instance with 53 qubits 20 cycles and sequence ABCDCDAB (There are several types of circuits considered in Google's experiments. The supremacy circuits used in the experiments are denoted as ABCDCDAB, and the simplified version used as verification of the experiments are denoted as EFGH), we choose $l=2^6$ and $L=2^{20}$, which leads to overall $2^{26}$ bitstring amplitudes obtained in one sparse-state contraction. The qubit ids of these 6 open qubits are $[11, 19, 28, 29, 37, 44]$ corresponding to the qubits layout in~\cite{pan_simulation_2022}. All $L$ bitstrings entries at the remaining 47 qubits subspace are generated uniformly at random.

\subsection{Slicing and contraction order}
If we adopt the above sampling scheme, the $L$ bitstrings which follow the Porter-Thomas distribution will be obtained, and their XEB fidelity will be 1, however, the computation requires huge computational resources. Since the samples obtained by Goggle has XEB around $0.002$, there is sill some space left to reduce the computational time by trading off the fidelity. There are basically two methods to achieve the trade-off.

The first method~\cite{markov2018quantum,villalonga2019flexible, huang2020classical} is to decrease the number of obtained bitstrings $L$. Intuitively, in general the noisy state of the quantum circuits can be expressed by
\begin{equation}
	\rho = f |\psi\rangle \langle \psi| + (1-f) \frac{1}{2^n}
\end{equation}
where $|\psi\rangle = U|\psi_0\rangle$ is the ideal state under no noise, $f$ is the fidelity. Then naturally, to obtain one million bitstrings with $0.002$ fidelity, one can generate $L=2000$ bitstrings with XEB $1$ then mix them to $998000$ uniformly and randomly generated bitstrings with $0$ fidelity.

The second way is to simplify the tensor network corresponding to the quantum circuit contraction. It has been shown that if we perform $k$ slicing to the tensor network to make $K$ copies of tensor network contractions, each slicing copy (or path in the picture of the Feynmann path integral) contributes $1/K$ to the final fidelity~\cite{markov2018quantum}. So summing over $f$ fraction of the overall slicing copies gives us $L$ bitstrings with fidelity $f$.

In the simulation of this work, we choose the second way to trade fidelity for the contraction complexity. Here we will firstly give a brief introduction to the slicing method in tensor network contractions, then discuss how we use the slicing method to decrease the overall contraction complexity of the simulation of the Sycamore circuits with $53$ qubits.

\begin{figure*}[htb]
	\centering
	\includegraphics[width=\columnwidth]{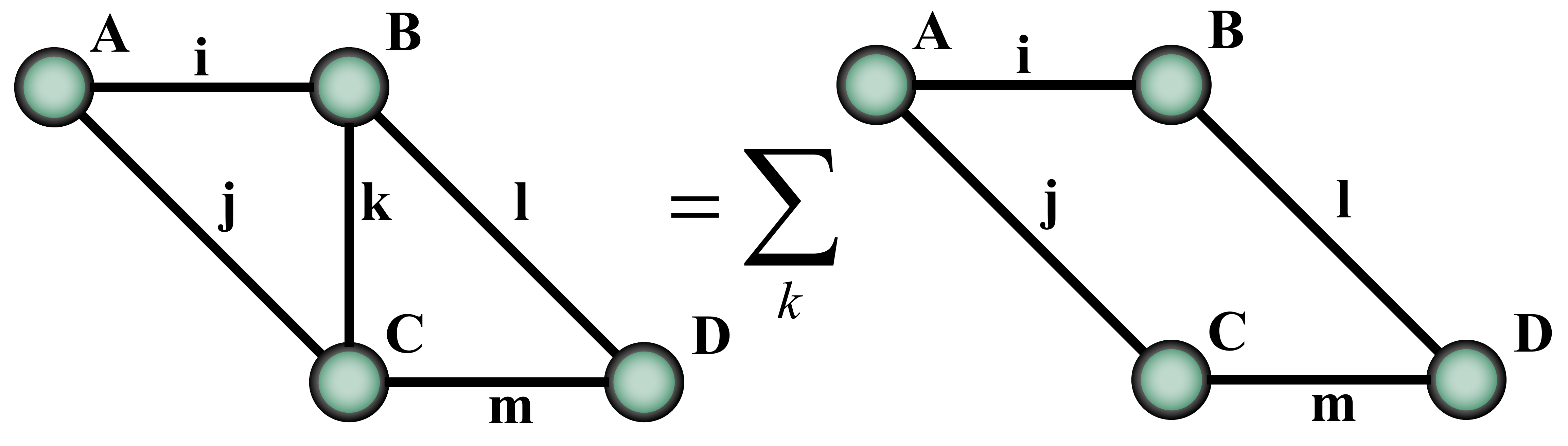}
	\caption{Illustration of the slicing. The index $k$ is sliced and the contraction of the original tensor network becomes summation of sub-tasks given different $k$ values. \label{fig:slicing}}
\end{figure*}

When the space complexity of a tensor network contraction is too large for the computational devices (i.e. GPU in this work), one needs to select some indices of the tensor network and fix them to some specific values in order to decrease the overall space complexity and fit the contraction into devices. Each choice of values of indices will lead to a sliced copy of the original tensor network, and contraction of the sliced copy is called a contraction sub-task. The summation of all sub-tasks results will return the contraction output of the original tensor network, as shown in Fig.~\ref{fig:slicing}.

There are several advantages of slicing in the tensor network corresponding to the random quantum circuits, especially ones use fSim gates as the two-qubit gates. The first one is that the sliced copies of the tensor network are all orthogonal to each other, leading to equal contributions to the fidelity. This allows us to sum only a fraction of all sub-tasks and allows an accurate estimate of the fidelity. The second advantage is that if we treat fSim gates as four-way tensors, slicing one index will make the slicing of companion index only cost a low fidelity decrease of overall contraction, due to the special low-rank structure of the fSim gate. A detailed explanation has been given in the main text about how to use fSim slicing to decrease the contraction complexity and how this kind of slicing will affect the final fidelity.

To obtain bitstring samples with XEB fidelity greater than $0.002$, we choose to sum $1/2^8$ of the overall sub-tasks, which leads to 8 breaking edges in the main text. Using the property of the fSim gates, we are able to remove $4$ two-qubit gates out of the original circuit. We can also do rank one approximations to companion edges of some slicing indices', this will give us a  $F_a$ factor of the overall fidelity. Thus the final estimate of the bitstring sample fidelity will be $F_{\text{estimate}} = F_a / 2^8$. 

The choice of slicing indices depends on the contraction order, a detailed explanation of how contraction order is determined has been given in the main text, in the following text we will explain how we choose the two-qubit gates to remove and how to remove the associated companion edges in the tensor network.

The tensor network corresponding to the Sycamore quantum circuit will be divided into the head and tail parts in order to determine the contraction order, as Fig.~\ref{fig:head} shows. Since the tail part involves all the final qubits, there will be at least 53 tensor indices at the interface between the head and tail part. In order to make the intermediate result of head contraction fit into a GPU with 32 GB of memory, 24 indices in the interface need to be sliced. The slicing edges in the interface are ``global slicing edges'' compared to "local slicing edges" which located within the head and tail parts. Each sub-task of a specific slicing-edge configuration gives $1/2^k$ of the overall contraction cost ($k$ is the number of slicing edges in the interface) while the local slicing edges do not affect the contraction cost of the counterpart. For the Sycamore circuits with $53$ qubits, we set all 8 breaking edges on the interface, which leads to 4 removed two-qubit gates. This increases the number of sub-tasks to $2^{16}$. After this, we also perform the rank one approximation to companion edges associated with 16 slicing edges on the interface, which also significantly decreases the overall contraction cost. 

We also need to put slicing inside the head tensor network and the tail tensor network, which is termed as ``local slicing edges'' because both the head and tail tensor network contraction require space complexity greater than $2^{30}$. The number of slicing edges of the head tensor network is set to 6 in the Sycamore circuits with $53$ qubits, and we do not perform rank one approximation to companion edges of the slicing edges. The number of slicing edges in the tail tensor network is set to 7, and we perform rank one approximation to companion edges of 5 slicing edges which are detected to decrease the contraction cost of the tail part.

\subsection{Computational cost}
In Tab.~\ref{table:data} we list the detailed data in the contractions of the Sycamore circuits with $n=53$ qubits and $m=20$ cycles.
\begin{table}[htb]
\begin{tabular}{|c|c|c|}
    \hline
    Data & Original & Branch merge \\
    \hline
    $T_c$ head one sub-task & $2.3816\times10^{13}$ & $6.967\times10^{13}$ \\
    \hline
    $T_c$ tail one sub-task & $2.9425\times10^{13}$ & $8.796\times10^{13}$ \\
    \hline
    Overall $T_c$ ($2^{16}$ sub-tasks) & $3.489\times10^{18}$ & $1.033\times10^{19}$ \\
    \hline
    Space complexity &  \multicolumn{2}{|c|}{$2^{30}$} \\
    \hline
    \# of slicing edges in $\ghead$ & \multicolumn{2}{|c|}{6} \\
    \hline
    \# of slicing edges in $\gtail$ & \multicolumn{2}{|c|}{7} \\
    \hline
    \# of slicing edges in the interface & \multicolumn{2}{|c|}{16} \\
    \hline
    \# of companion edges in $\ghead$ & \multicolumn{2}{|c|}{0} \\
    \hline
    \# of companion edges in $\gtail$ & \multicolumn{2}{|c|}{5} \\
    \hline
    \# of companion edges in the interface & \multicolumn{2}{|c|}{16} \\
    \hline
    Fidelity of rank one approximation & \multicolumn{2}{|c|}{0.9564714760983217} \\
    \hline
    GPU efficiency head & - & 31.76\% \\
    \hline
    GPU efficiency tail & - & 14.27\%\\
    \hline
    Overall efficiency & - & 18.85\%\\
    \hline
\end{tabular}
\caption{Detailed data in the contractions of the Sycamore circuits with $n=53$ qubits and $m=20$ cycles. $T_c$ represents the time complexity. The companion edges in the fSim slicing can be used to decrease the contraction complexity. In the simulation, we include no companion edges in the head, $5$ out of $7$ in the tail, and all $16$ in the interface. Some companion edges are excluded because they do not contribute much to the overall complexity and including them will result in a drop in fidelity. The calculated fidelity comes from $21$ companion edges, each of them contributes a factor $(\sin^2{\theta}+1)/2$ to the fidelity ($\theta$ is the parameter of the associated fSim gate). The GPU efficiencies are calculated by $8\cdot T_c / (P\cdot t)$ where $P$ is the single-precision performance of GPU, $t$ is the computation time and $8$ is the factor of matrix multiplication for complex number ($8\cdot T_c$ denotes the overall floating-point operations of the contraction).\label{table:data}}
\end{table}

\section{The singular values of the sliced fSim gate}
Here we give a detailed calculation of the singular values of the sliced fSim gate.
The matrix form of the fSim gate is
\begin{equation}
    \label{eq:fsim}
    \text{fSim}(\theta, \phi) = 
    \begin{bmatrix}
        1 & 0 & 0 & 0 \\
        0 & \cos{\theta} & -i\sin{\theta} & 0 \\
        0 & -i\sin{\theta} & \cos{\theta} & 0 \\
        0 & 0 & 0 & e^{-i\phi}
    \end{bmatrix}\;.
\end{equation}
with $\theta\approx \frac{\pi}{2}$. In the above matrix, the first dimension corresponds to the dimension of two input qubits and the first dimension corresponds to the dimension of two output qubits of the fSim gate.
We consider two kinds of slicing to the fSim gate.
The first slicing is to pin two input indices $\alpha$ and $\beta$. I.e., two input edges of the $\fsim$ gate are cut, as shown in the top panel of Fig.~\ref{fig:fsim}. This is equivalent to applying two Pauli errors gate as 
$A = \left(\left[\begin{matrix}1&0\\0&0\end{matrix}\right]\otimes \left[\begin{matrix}1&0\\0&0\end{matrix}\right]\right)\cdot\fsim(\theta,\phi),$.
    The resulting matrix is $B=\left[\begin{matrix}1&0\\0&0\end{matrix}\right]$, with the input dimension and output dimension corresponding to the first and second output qubits. Notice that $B$ can be evaluated as $B=\left( \begin{matrix} 1\\0\end{matrix}\right)\otimes \left( \begin{matrix} 1\\0\end{matrix}\right)$, which also pins the output qubits. So in the picture of tensor network diagram, the entire fSim gate is compltely replaced by two $(1,0)$ vectors, which reduces the computational complexity heavily. Notice that the rank-one structure is exact so the replacing step does not decrease fidelity.\\

        The second slicing situation we consider is pinning (i.e. fixing the index of) one input edge of the fSim gate, as shown in as illustrated in Fig.~\ref{fig:fsim} bottom (e.g. the top left edge $\gamma$ of tensor $D$ is cut). This gives a three-way tensor $F$. Although the decompositional rank of $F$ on the bottom right index $\omega$ is $2$, the corresponding singular values might be are heavily imbalanced for the Sycamore circuits with $\theta\approx {\pi/2}$ so that we can perform accurate low-rank approximations.
        Now let us compute explicitly the eigenvalues of the resulting matrix by reshaping $E$ for all $4$ possible configurations of slicing: pinning the first input qubit to $0$; pinning the first input qubit to $1$; pinning the second input qubit to $0$; and pinning the second input qubit to $1$.

                \begin{enumerate}
            \item Pinning the first input qubit to $0$. The fSim gate becomes 
                $$F_{0***} = \left( \begin{matrix} 1&0&0&0\\0&\cos\theta&-i\sin\theta&0\end{matrix}\right) $$
                    where $F_{0***}$ has $4$ indices, corresponding to the first input qubit, the second input qubit, the first output qubit, and the second output qubit respectively. $0$ means that the index is pinned to the first value, and $*$ means that the index can take $2$ values and hence is not sliced.
                Reshaping the above matrix by grouping the first three dimensions of the gate, we have 
                $$F_{(0**)*} = \left( \begin{matrix} 1&0\\0&0\\0&\cos\theta\\-i\sin\theta&0\end{matrix}\right) $$
                    and we can see that the squared singular values are given by
                $$F_{(0**)*}^\dagger F_{(0**)*} = \left( \begin{matrix} 1+\sin^2\theta &0\\0&\cos^2\theta\end{matrix}\right) $$

                \item 
Pinning the second input qubit to $0$. The fSim gate becomes 
                $$F_{*0**} = \left( \begin{matrix} 1&0&0&0\\0&-i\sin\theta&\cos\theta&0\end{matrix}\right) $$
                    By swapping the third and the fourth dimension, we have
                    $$F_{*0* *} = \left( \begin{matrix} 1&0&0&0\\0&\cos\theta&-i\sin\theta&0\end{matrix}\right) $$
                Reshaping the above matrix by grouping the first three dimensions of the gate, we have 
                $$F_{(*0*)*} = \left( \begin{matrix} 1&0\\0&0\\0&\cos\theta\\-i\sin\theta&0\end{matrix}\right) $$
                    and we can see that the squared singular values are given by
                $$F_{(*0*)*}^\dagger F_{(*0*)*} = \left( \begin{matrix} 1+\sin^2\theta &0\\0&\cos^2\theta\end{matrix}\right) $$

                \item 
Pinning the second input qubit to $0$. The fSim gate becomes 
$$F_{1***} = \left( \begin{matrix} 0&-i\sin\theta&\cos\theta&0\\0&0&0&e^{-i\phi}\end{matrix}\right) $$
                Reshaping the above matrix by grouping the first three dimensions of the gate, we have 
                $$F_{(1**)*} = \left( \begin{matrix} 0&-i\sin\theta\\\cos\theta&0\\0&0\\0&e^{-i\phi}\end{matrix}\right) $$
                    and we can see that the squared singular values are given by
                $$F_{(1**)*}^\dagger F_{(1**)*} = \left( \begin{matrix} \cos^2\theta &0\\0&1+\sin^2\theta\end{matrix}\right) $$

                \item 
Pinning the second input qubit to $1$. The fSim gate becomes 
$$F_{*1**} = \left( \begin{matrix} 0&\cos\theta&-i\sin\theta&0\\0&0&0&e^{-i\phi}\end{matrix}\right) $$
                    By swapping the third and the fourth dimension, we have
$$F_{*1**} = \left( \begin{matrix} 0&-i\sin\theta&\cos\theta&0\\0&0&0&e^{-i\phi}\end{matrix}\right) $$
                Reshaping the above matrix by grouping the first three dimensions of the gate, we have 
                $$F_{(*1*)*} = \left( \begin{matrix} 0&-i\sin\theta\\\cos\theta&0\\0&0\\0&e^{-i\phi}\end{matrix}\right) $$
                    and we can see that the squared singular values are given by
                $$F_{(*1*)*}^\dagger F_{(*1*)*} = \left( \begin{matrix} \cos^2\theta &0\\0&1+\sin^2\theta\end{matrix}\right) $$

        \end{enumerate}
Here we can see that although the decompositional rank of $F$ on the bottom right index $\omega$ is $2$, the corresponding squared singular values, $\left(\sin^2(\theta)+1,\cos^2(\theta)\right )$, are heavily imbalanced in the Sycamore circuits with $\theta\approx {\pi/2}$. So we can do a rank-one approximation by dropping the singular vectors corresponding to the squared singular value $\cos^2(\theta)$. This rank-one approximation decreases the fidelity approximately by a factor $(\sin^2(\theta)+1)/2$, while effectively break another edge $\omega$, which we term as the \textit{companion edge} in the tensor network. 

\begin{figure}[h]
    \centering
    \includegraphics[width=0.8\columnwidth]{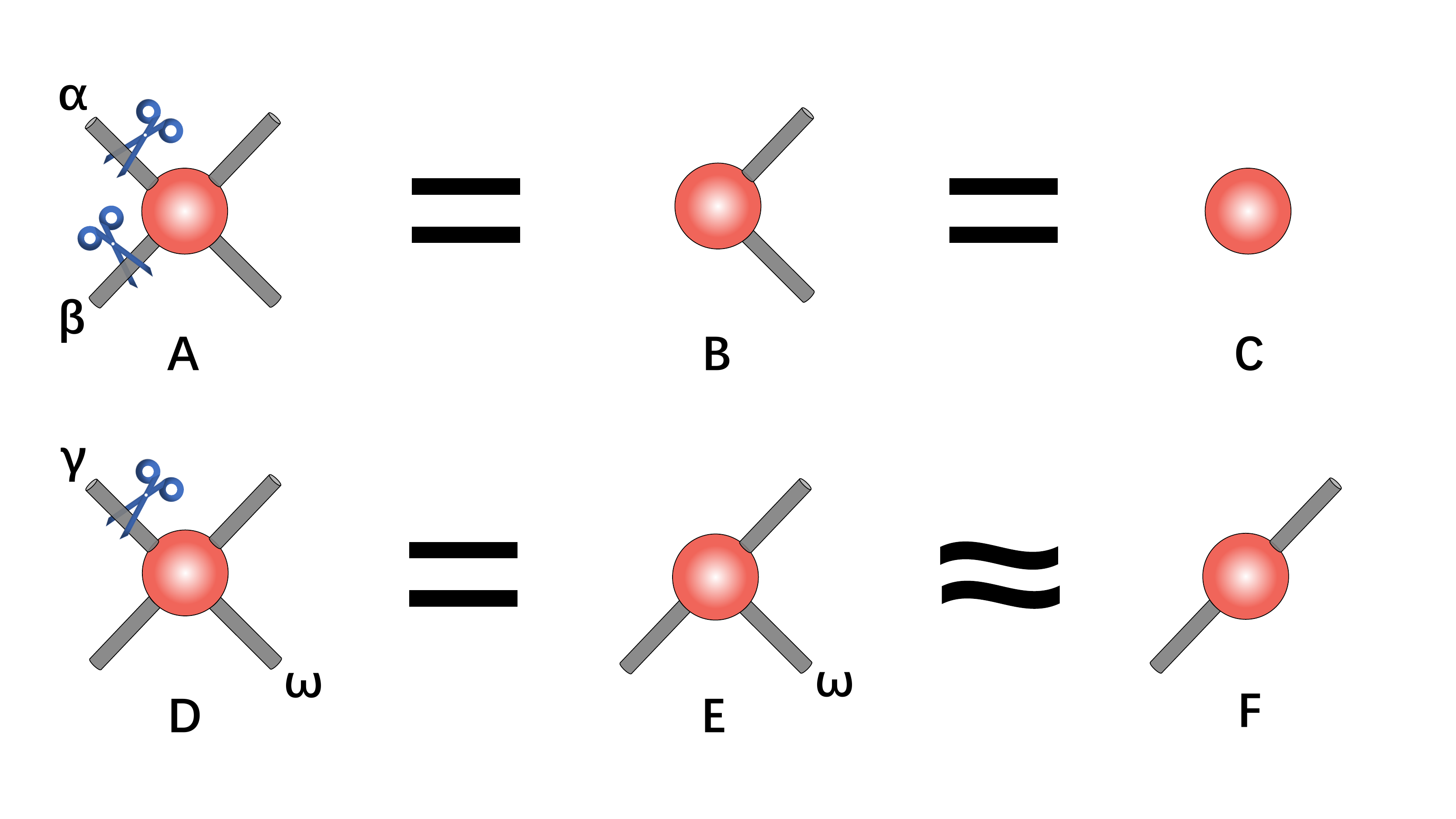}
    \caption{Two situations that we can explore the low-rank structures. (Top:) When two indices $\alpha,\beta$ are pinned to $0$ for the fSim gate. The result is a rank-one matrix $B$, effectively equals to a scalar $c=1$ in tensor network. (Bottom:) When one index $\gamma$ of a fSim gate is pinned, the resulting tensor $E$ has decompositional rank $2$ but with imbalanced singular values $\left(\sqrt{\sin^2(\theta)+1},\cos(\theta)\right )$ with $\theta\approx \pi/2$.
    \label{fig:fsim}}
\end{figure}

\section{Validation of our method using smaller Sycamore circuits}
The estimated fidelity $F_{\mathrm{estimate}}\approx 0.0037$ of one million samples generated using our method does not require defining a proxy of fidelity, such as the XEB~\cite{arute2019quantum}. But one can use XEB to verify the fidelity value that we claim, e.g. by utilizing the verification method reported in~\cite{kalachev_recursive_2021}.
However verifying one million samples for $n=53$ qubits, $m=20$ cycles cost a huge amount of computational resources that we can not afford. So in this section, we show validation of our approximate sampling method using smaller Sycamore circuits where we can compute the exact amplitudes of the original circuit, and easily verify the fidelity of $\hpsi$ and XEB of generated samples.

\begin{figure}[h]
    \centering
    \includegraphics[width=0.45\columnwidth]{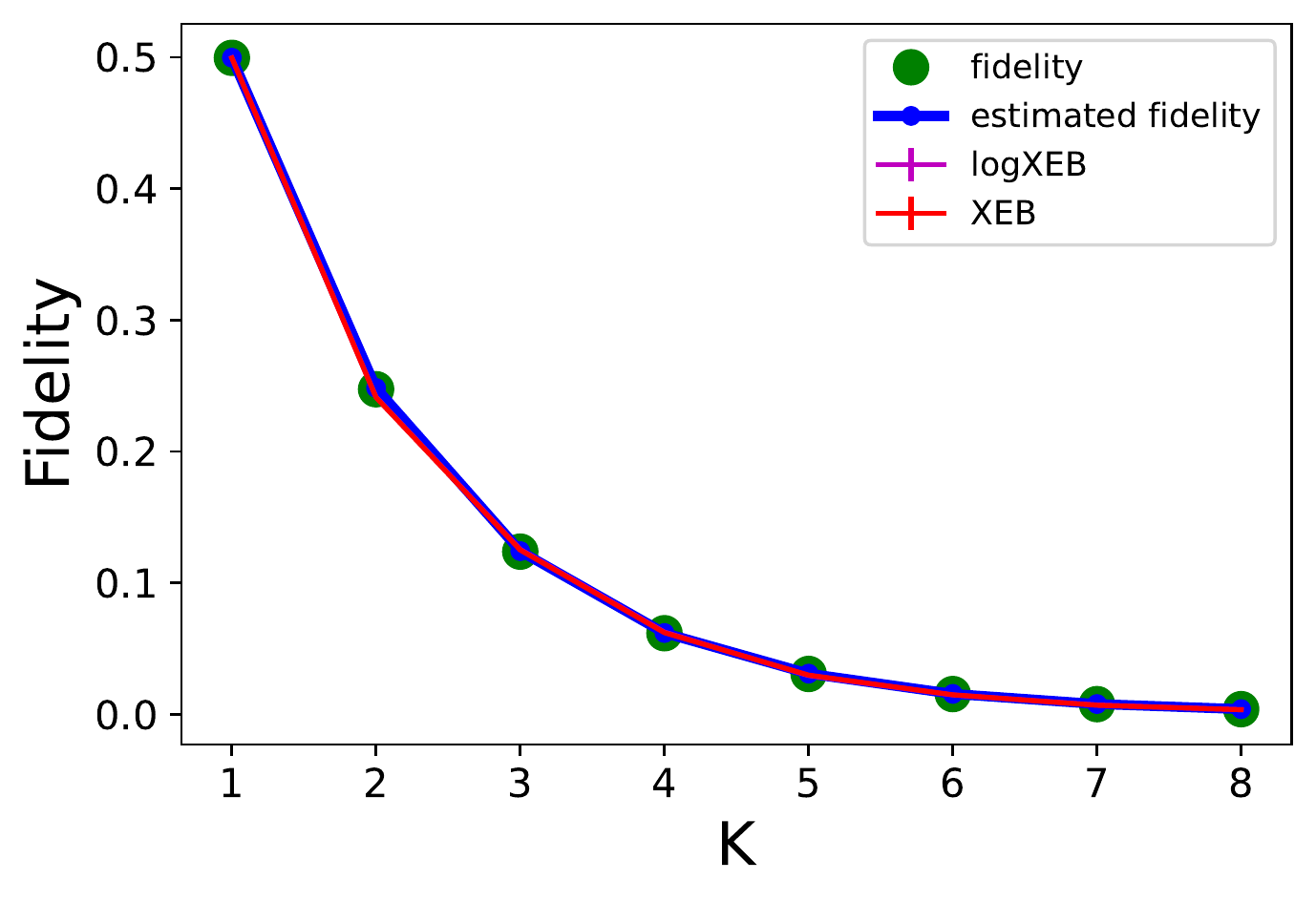}
    \includegraphics[width=0.47\columnwidth]{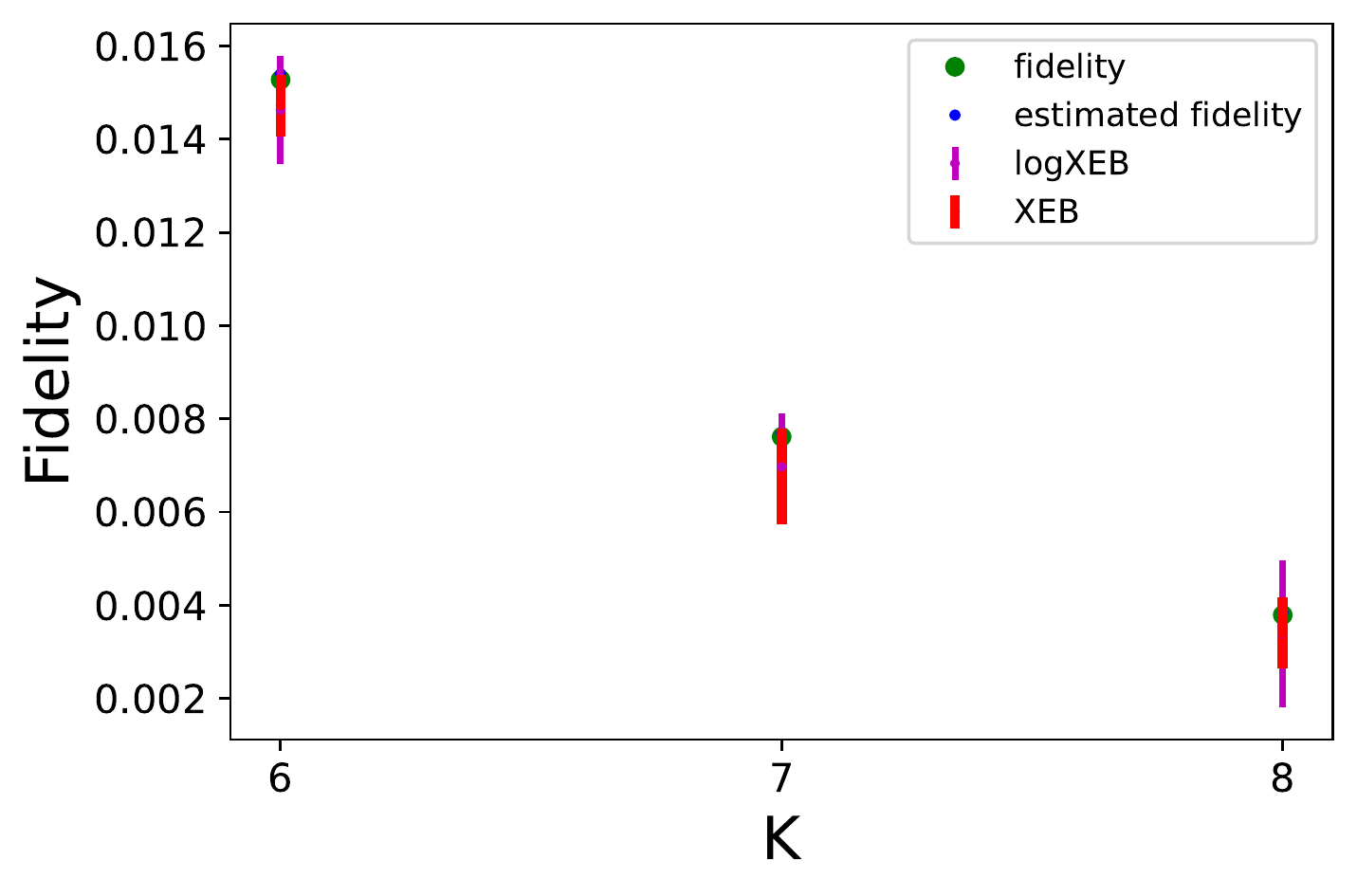}
    \caption{\textit{Left:} Comparison between the exact fidelity of approximate state $\widehat \psi_K(\s)$, estimated fidelity (computed using the method described in the main text), the logarithmic XEB~\cite{arute2019quantum} and the XEB value of $2^{20}$ bitstrings sampled from $\widehat P(\s)=|\widehat\psi_K(\s)|^2$. The approximate state $\widehat \psi_K$ is obtained by breaking $K$ edges in the tensor network, for the Sycamore circuits with $n=30$ qubits, $m=14$ cycles, and EFGH sequence. Each data point is averaged over $15$ independent sets of samples of size $2^{20}$. The errorbar is much smaller than the symbol size. \textit{Right:} The enlarged figure of the left panel focusing on $K=6, 7, $ and $8$. \label{fig:verification}}
\end{figure}

We choose a Sycamore circuit with $n=30$ qubits, $m=14$ cycles, and sequence EFGH, compute and store the exact final state $\psi$. 
Then we remove $K$ edges associated with $\fsim$ gates together with $K$ companion edges in the middle of the circuit, mimicking what we did for the circuits with $n=53$ qubits and $m=20$ cycles. Then we compute the corresponding $\psi_K$ exactly by evaluating the state vector and compute the fidelity $F$.
In Fig.~\ref{fig:verification}, we compare the true fidelity $F$ and the estimated fidelity $ F_{\mathrm{estimate}}$ (using the method described in the main text), we can see from the figure that they coincide very well. 
From the probability distribution associated with the approximate state vector $\hp=|\hpsi|^2$, we can obtain a set of independent samples using the reject sampling method described in the main text. Analogous to the sampling procedure we performed for the circuits with $n=53$ qubits and $m=20$ cycles. On the circuit with $n=30$ qubits and $m=14$ cycles we also computed approximate probabilities $\{\hp (\s_i)\}$ for $2^{26}$ bitstrings $\{\s_i\}$ which are grouped into $2^{20}$ groups. Then we sample one bitstring from each group using the rejection sampling with Markov chains and finally produce $2^{20}$ uncorrelated bitstrings $\{s_i|i=1,2,\cdots 2^{20}\}$. Since we have stored the exact state vector, we can compute exact probabilities $P(\s_i)$ for each sample and evaluate the XEB for the samples using
$$
F_{\mathrm{XEB}} = \frac{2^{30}}{2^{20}}\sum_{i=1}^{2^{20}}P(s_i)-1.
$$
The sampling and XEB calculation are repeated for $15$ times and the average XEB values are shown in Fig.~\ref{fig:verification} and compared with the estimated fidelity $F_{\mathrm{estimate}}$ as well as the true fidelity $F$, which we can see all of them agree to each other. The right panel of Fig.~\ref{fig:verification} shows more clearly that the error bars are small and the XEB values are significantly greater than Google's value of $0.002$.

\begin{figure}[h]
    \centering
    \includegraphics[width=0.6\columnwidth]{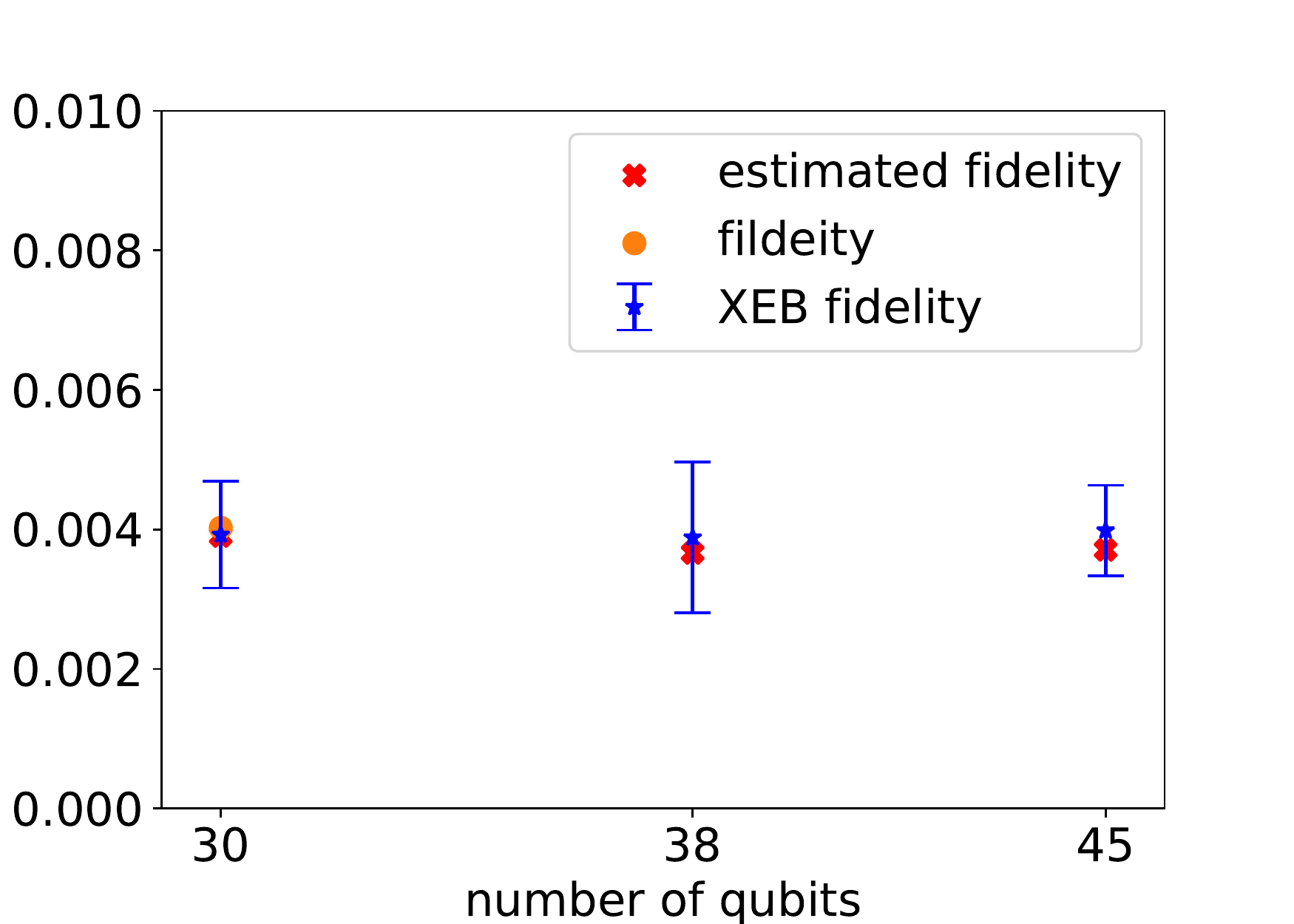}
    \caption{\textit{Left:} Comparison between the XEB of $2^{20}$ samples, estimated fidelity (computed using the method described in the main text), and the exact fidelity of approximate state $\widehat \psi_K(\s)$, for the Sycamore quantum circuits with $n=30, 38, $ and $45$ qubits, $m=14$ cycles, and EFGH sequence. Each data point is averaged over $30$ independent sets of $2^{20}$ bitstring samples.\label{fig:errorbar}}
\end{figure}

In Fig.~\ref{fig:errorbar}, we show comparisons between the estimated fidelity and the computed XEB for circuits with $K=8$ cuts, different number of qubits, and errorbars obtained over different sets of samples. We can see that the fidelity of the sparse state and the XEB of obtained samples are stably consistent with an increasing the number of qubits.

\begin{figure}[h]
    \centering
    \includegraphics[width=0.6\columnwidth]{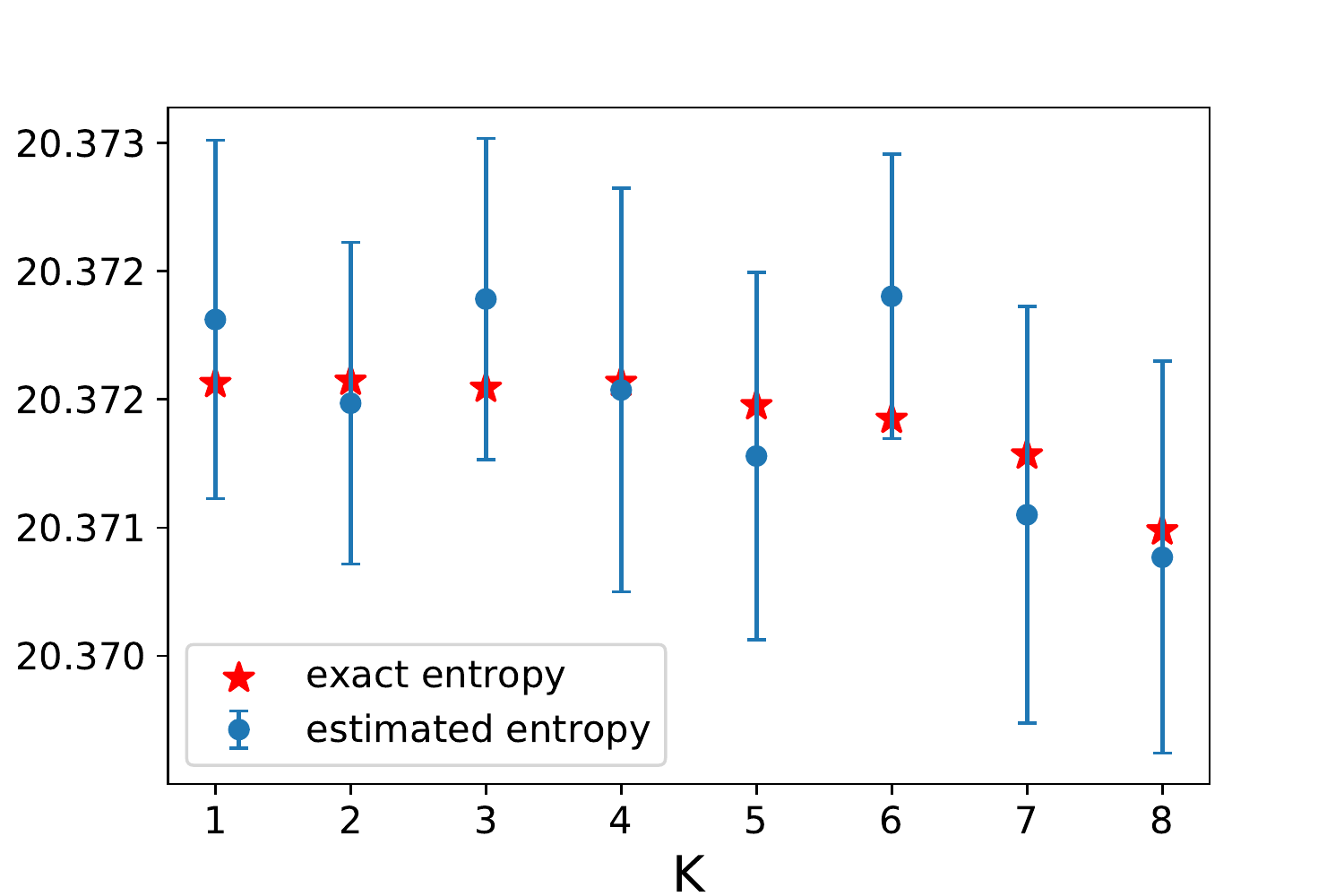}
    \caption{The exact entropy of the approximate state distribution of $|\widehat \psi_{K}(\s)|^2$ compared with the entropy estimated using $2^{20}$ uncorrelated samples generated using the sampling method proposed in the main text, for $K$ cuts in the Sycamore circuits with $n=30$ qubits, $m=14$ cycles, and EFGH sequence. Each data point is averaged over $15$ independent sets of samples of size $2^{20}$.\label{fig:entropy}}
\end{figure}

In Fig.~\ref{fig:entropy}, we compare the entropy estimated using $2^{20}$ samples and the entropy of the distribution corresponding to the sparse state for Sycamore circuits with $n=30$ qubits. We can see that the entropy estimated using the samples coincides very well with the entropy of the distribution that the samples came from. This indicates that the sampling algorithm indeed satisfies similar properties of the underline distribution.

\begin{figure}[h]
    \centering
    \includegraphics[width=0.6\columnwidth]{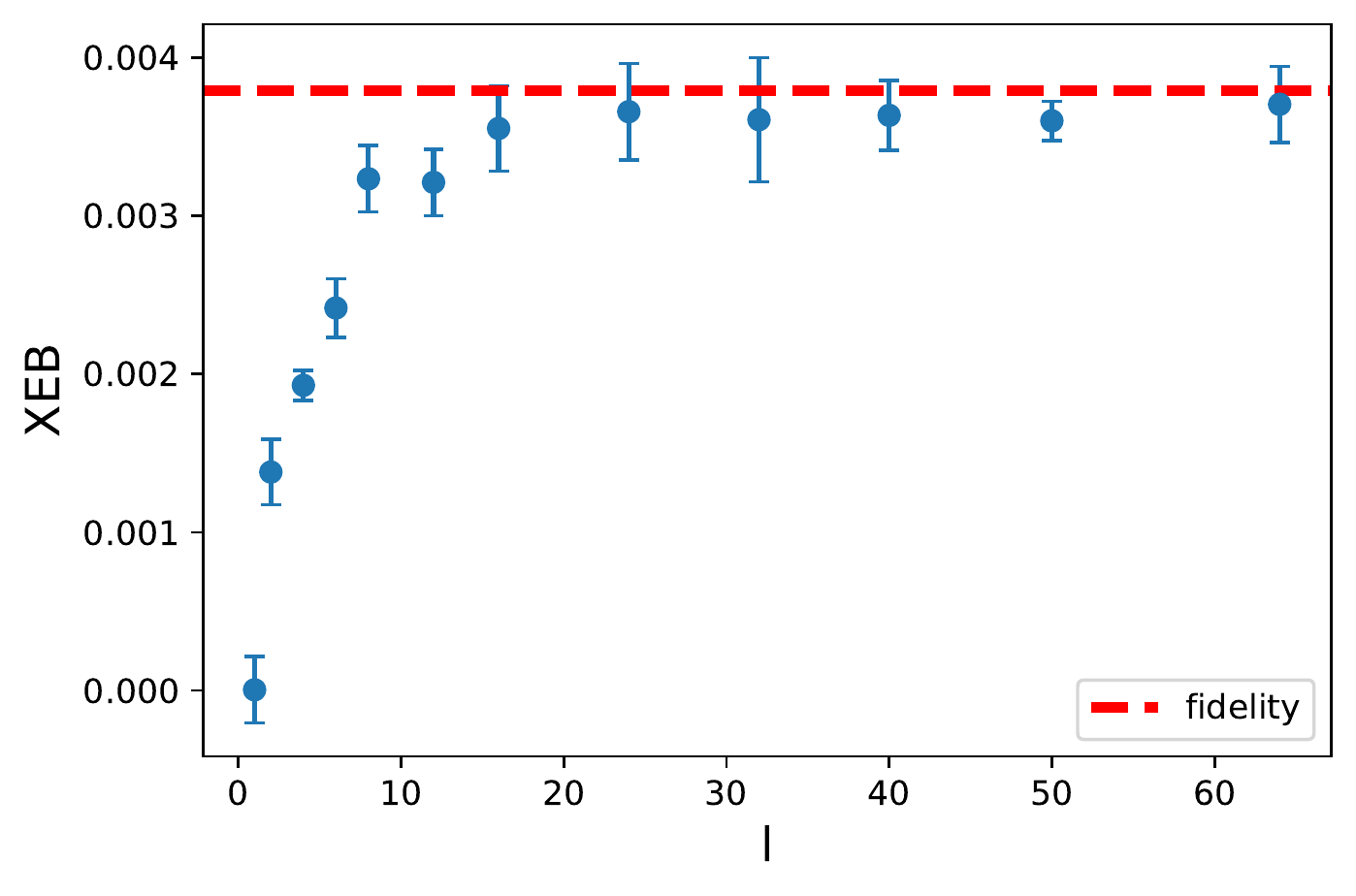}
    \caption{ The XEB computed using $2^{20}$ uncorrelated samples as a function of a number of group size $l$ for the Sycamore circuits with $n=30$ qubits, $m=14$ cycles, and EFGH sequence, and with $K=8$ cuts. Each data point is averaged over $15$ independent sets of samples of size $2^{20}$ The horizontal dashed line is the exact fidelity of the approximate state $\widehat \psi_{K=8}(\s)$. \label{fig:xebl}}
\end{figure}

In Fig.~\ref{fig:xebl} we evaluated the XEB of samples as a function of group size $l$ (see main text), and compare the XEB values to the exact fidelity of the approximate state $\widehat \psi_{K=8}(\s)$, with $K=8$ cuts for the Sycamore circuits with $n=30$ qubits, $m=14$ cycles, and EFGH sequence. From the figure, we can see that $l = 64$ is sufficient for the sampling tas.

%We have also checked the XEB value of samples generated using the frugal sampling method, which also agrees with the XEB of our reject sampling and fidelities.

%\begin{figure}[h]
%    \centering
%    \includegraphics[width=0.9\columnwidth]{verify_n30_m14_20samples.pdf}
%    \caption{Comparison between the exact fidelity of approximate state $\widehat \psi_K(\s)$, estimated fidelity (computed using the method described in described in the main text), and the XEB value of $2^{20}$ bitstrings sampled from $\widehat P(\s)=|\widehat\psi_K(\s)|^2$. The approximate state $\widehat \psi_K$ is obtained by breaking $K$ edges in the tensor network, for the Sycamore circuits with $n=30$ qubits, $m=14$ cycles, and EFGH sequence.\label{fig:verification}}
%\end{figure}

\begin{figure*}[htb]
	\centering
	\includegraphics[width=\columnwidth]{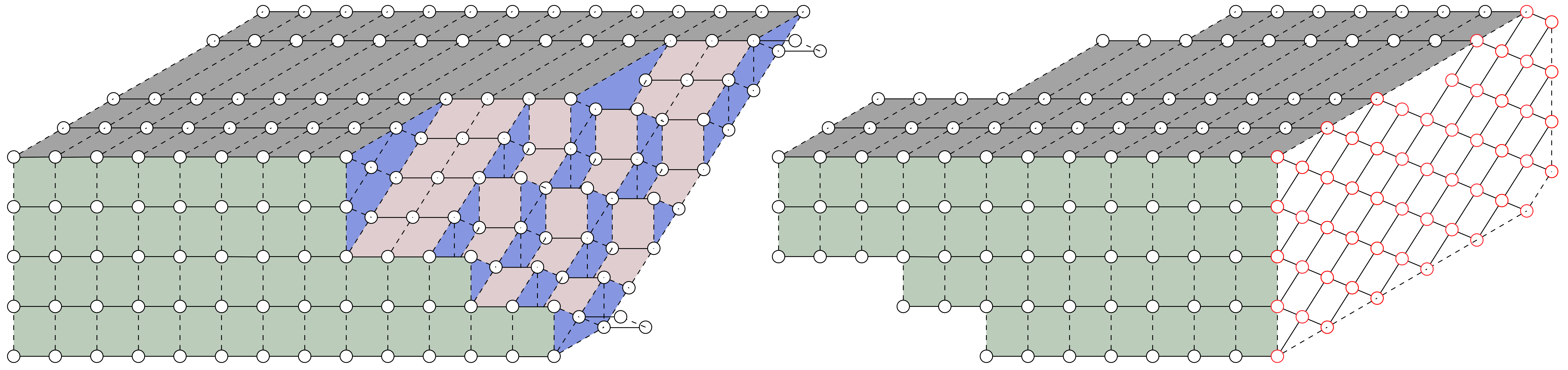}
	\caption{The 3-dimensional tensor network $\hg$ (corresponding to the Sycamore circuit of $n=53$ qubits, $m=20$ cycles) is split into two parts, $\ghead$ (left) and $\gtail$ (right). \label{fig:head}}
\end{figure*}

\section{Contraction order}\label{sec:order}
Here we list the contraction order that has been used in our simulation of the Sycamore circuit with $n=53$ qubits, $m=20$ cycles, containing the Zig-Zag ordering for the tail part of the tensor network contraction.

\begin{figure*}[h]
	\centering
	\includegraphics[width=0.9\columnwidth]{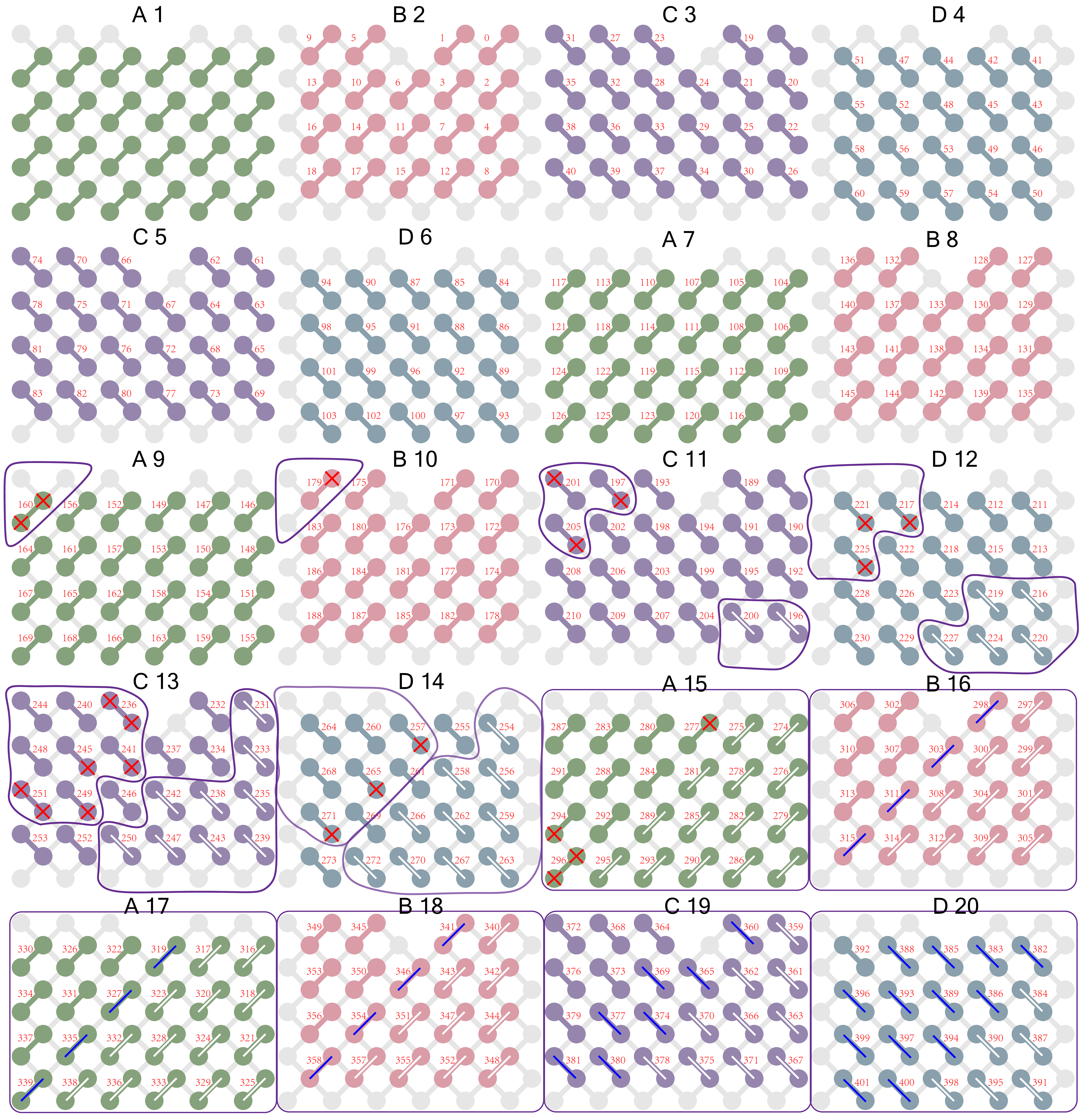}
	\caption{The head-tail splitting and details about the contraction scheme in the view of two-dimensional qubit layout. The two-qubit gates of all $20$ cycles with IDs in our actual contraction code and their sequence patterns in the Sycamore circuit are listed. Note that some gates are contracted out during the simplification stage and thus have no corresponding IDs. The gates wrapped by purple lines belong to $\gtail$ while others belong to $\ghead$. The red crosses over gates represent the slicing indices that connect the former layer to gates in the current layer. Notice that the four gates with two slicing crosses are the miss blocks. The white and blue lines over the gates denote the zig-zag contraction order of the tail part, white for the first zig and blue for the first zag. The following order can be easily repeated accordingly.\label{fig:grid}}
\end{figure*}

After contracting two tensors at location $i$ and $j$, the resulting new tensor will be put back to location $i$. In the following order, the contractions from (47, 70) to (192, 234) belong to the head part, and the rest belong to the tail part. The head-tail spliting and the IDs of tensors are explicitly illustrated in Fig.~\ref{fig:head} and Fig.~\ref{fig:grid}.
In the figure, contractions involving the C11 to D20 belong to the tail part, where the contraction order is the zig-zag order manually designed by us. The zig-zag order starts from the left boundary (which separates the head and the tail networks) to the right boundary (the sparse state), then back to the left boundary. In the figure, the contraction sequence is illustrated using the white and blue lines over the gates, white for the first zig and blue for the first zag. The following order can be easily repeated accordingly.

[(47, 70), (27, 47), (117, 136), (27, 117), (35, 55), (78, 35), (9, 78), (27, 9), (5, 13), (31, 74), (51, 31), (94, 51), (75, 94), (32, 52), (75, 32), (79, 98), (95, 118), (175, 193), (156, 175), (132, 156), (183, 132), (121, 164), (140, 121), (183, 140), (113, 183), (137, 152), (110, 137), (180, 202), (161, 180), (110, 161), (44, 66), (23, 44), (67, 87), (23, 67), (38, 81), (58, 38), (24, 45), (186, 208), (124, 143), (167, 124), (60, 83), (103, 60), (40, 103), (145, 169), (126, 145), (57, 100), (80, 57), (77, 97), (54, 77), (253, 273), (230, 253), (210, 230), (125, 144), (168, 125), (210, 168), (73, 93), (50, 73), (116, 50), (26, 69), (46, 26), (209, 228), (142, 166), (123, 142), (229, 252), (184, 206), (207, 226), (22, 43), (65, 22), (21, 65), (42, 64), (85, 42), (21, 85), (86, 21), (1, 86), (0, 61), (41, 62), (19, 41), (128, 171), (149, 128), (176, 198), (199, 218), (147, 170), (127, 147), (131, 148), (106, 131), (222, 246), (203, 222), (223, 203), (120, 163), (139, 120), (182, 139), (154, 182), (214, 237), (194, 214), (215, 194), (189, 211), (232, 189), (212, 255), (234, 212), (191, 213), (234, 191), (159, 178), (27, 5), (27, 75), (27, 79), (27, 95), (76, 27), (113, 110), (90, 113), (76, 90), (71, 76), (71, 23), (28, 10), (28, 58), (28, 71), (56, 16), (56, 36), (56, 28), (14, 24), (14, 6), (14, 56), (101, 186), (101, 167), (101, 14), (18, 82), (18, 40), (101, 18), (33, 53), (33, 39), (101, 33), (17, 48), (17, 37), (101, 17), (11, 59), (11, 29), (101, 11), (15, 25), (15, 7), (101, 15), (34, 126), (34, 122), (101, 34), (80, 12), (80, 99), (101, 80), (4, 102), (4, 54), (101, 4), (96, 210), (96, 119), (101, 96), (188, 141), (188, 165), (101, 188), (114, 72), (114, 91), (101, 114), (49, 30), (49, 3), (101, 49), (8, 116), (8, 46), (101, 8), (68, 155), (68, 88), (101, 68), (111, 138), (111, 92), (101, 111), (187, 162), (187, 209), (101, 187), (115, 123), (115, 229), (101, 115), (184, 157), (184, 89), (101, 184), (133, 107), (133, 207), (101, 133), (185, 158), (185, 134), (101, 185), (108, 1), (101, 108), (109, 135), (109, 2), (101, 109), (0, 19), (0, 20), (101, 0), (63, 149), (101, 63), (84, 105), (84, 176), (101, 84), (104, 129), (104, 130), (101, 104), (150, 177), (150, 153), (101, 150), (181, 199), (181, 223), (101, 181), (127, 112), (101, 127), (106, 173), (101, 106), (146, 172), (146, 151), (101, 146), (190, 232), (190, 204), (101, 190), (174, 195), (192, 174), (192, 154), (192, 101), (159, 215), (192, 159), (192, 234), (359, 402), (359, 403), (360, 404), (382, 405), (382, 406), (361, 407), (383, 408), (383, 409), (384, 410), (384, 411), (363, 412), (385, 414), (385, 415), (386, 416), (386, 417), (387, 418), (387, 419), (367, 420), (388, 422), (388, 423), (389, 424), (389, 425), (390, 426), (390, 427), (391, 428), (391, 429), (392, 432), (393, 433), (393, 434), (394, 435), (394, 436), (395, 437), (395, 438), (396, 440), (396, 441), (397, 442), (397, 443), (398, 444), (398, 445), (399, 447), (399, 448), (400, 449), (400, 450), (381, 451), (401, 452), (401, 453), (339, 454), (179, 160), (197, 179), (217, 197), (241, 217), (261, 241), (225, 205), (249, 225), (269, 249), (245, 221), (265, 245), (284, 265), (271, 251), (292, 271), (257, 236), (277, 257), (196, 216), (196, 200), (192, 196), (227, 250), (227, 224), (192, 227), (219, 220), (219, 242), (192, 219), (231, 233), (231, 254), (192, 231), (235, 238), (235, 256), (192, 235), (239, 243), (239, 259), (192, 239), (247, 258), (247, 266), (192, 247), (262, 282), (262, 263), (192, 262), (267, 270), (267, 286), (192, 267), (272, 274), (272, 293), (192, 272), (275, 276), (275, 297), (192, 275), (278, 279), (278, 299), (192, 278), (285, 290), (285, 304), (192, 285), (301, 305), (301, 318), (192, 301), (309, 312), (309, 324), (192, 309), (320, 321), (320, 361), (192, 320), (316, 325), (316, 342), (192, 316), (328, 329), (328, 347), (192, 328), (333, 336), (333, 352), (192, 333), (344, 348), (344, 355), (192, 344), (366, 363), (367, 366), (192, 367), (375, 371), (387, 375), (192, 387), (261, 281), (261, 300), (261, 317), (261, 391), (192, 261), (269, 289), (269, 308), (269, 323), (269, 395), (192, 269), (340, 343), (340, 362), (340, 359), (192, 340), (295, 314), (295, 332), (295, 351), (295, 384), (192, 295), (338, 357), (338, 370), (338, 390), (192, 338), (378, 398), (382, 360), (378, 382), (378, 383), (192, 378), (386, 389), (386, 394), (386, 397), (386, 365), (192, 386), (400, 401), (400, 380), (400, 374), (192, 400), (385, 393), (385, 399), (341, 298), (341, 277), (341, 319), (341, 346), (385, 341), (192, 385), (369, 388), (303, 327), (303, 354), (369, 303), (192, 369), (377, 396), (284, 311), (284, 335), (284, 377), (192, 284), (381, 339), (381, 358), (381, 315), (381, 292), (381, 296), (381, 294), (381, 337), (381, 356), (381, 379), (192, 381), (313, 331), (313, 350), (313, 373), (313, 288), (313, 307), (313, 322), (313, 280), (313, 364), (313, 260), (313, 268), (313, 240), (192, 313), (248, 291), (248, 334), (248, 310), (248, 353), (248, 283), (192, 248), (326, 302), (326, 345), (326, 264), (326, 287), (326, 306), (192, 326), (392, 372), (392, 244), (392, 201), (392, 349), (392, 330), (392, 368), (392, 376), (192, 392)]. 

\end{document}